\documentclass{aa}  

\usepackage{xcolor}
\usepackage{graphicx}
\usepackage{txfonts}
\usepackage{lipsum}
\usepackage{subcaption}         
\usepackage{lscape}             
\usepackage{placeins}           
\usepackage[normalem]{ulem}
\usepackage{hyperref}
\usepackage{orcidlink}

\begin{document}

\title{A multi-fluid approach for polydisperse pebble accretion}
   
\subtitle{From particles to fluids, establishing the multi-fluid framework} 

\author{T. J. Konijn\orcidlink{0000-0003-4376-6287} 
        \and 
        S.-J. Paardekooper\orcidlink{0000-0002-8378-7608}
        }

   \institute{Planetary Exploration Group, Faculty of Aerospace Engineering, Delft University of Technology, Kluyverweg 1, 2629 HS Delft, The Netherlands \\
   email: \href{mailto:t.j.konijn@tudelft.nl}{T.J.Konijn@tudelft.nl}; \href{mailto:s.paardekooper.nl}{S.Paardekooper@tudelft.nl}}

   \date{Received December 5, 2025 / Accepted 22 April 2026}
   
\abstract
     {Pebble accretion offers an efficient pathway to form planets, driven by a constant supply of inward drifting mass and an accretion efficiency enhanced by gas drag. While most studies assume a single pebble size (monodisperse), real discs contain a range of sizes (polydisperse) that drift, interact, and accrete at different rates.}
  {We aim to model polydisperse pebble accretion with a fluid approach, validating the method and exploring how gas disc evolution, solid-to-gas back-reaction, and a polydisperse size distribution affect growth.}%
   {We used FARGO3D, modified to allow pebble accretion, to run 2D hydrodynamic simulations in a global disc with multiple pebble species representing an underlying continuous pebble size distribution.}
   {With our multi-fluid approach, we find values for pebble accretion efficiency consistent with earlier studies for a static gas disc. This confirms that our approach gives an accurate representation of pebble accretion. Evolving the gas disc, we find lower efficiencies compared to an unperturbed gas disc for high Stokes numbers ($\gtrsim 0.3$) and higher efficiencies for smaller Stokes numbers ($\lesssim0.3$). This effect increases for higher planet masses. The accretion rate is mostly dominated by the highest Stokes numbers in our parameter study ($\mathrm{St}\in[10^{-2},10^0]$). The ratio we find between the polydisperse and monodisperse pebble accretion rates is higher than previous estimations.} 
  {We constructed a multi-fluid model framework capable of accurately simulating polydisperse pebble accretion consistent with previous studies. This framework offers advantages for simulating higher planet masses and for modelling multiple pebble species coupled to the gas. We find that  the protoplanet's perturbation of the gas-disc lowers the accretion rate when assuming an MRN-distribution of solids.}%


   \maketitle
   \nolinenumbers


\section{Introduction}
\label{sec:introduction}

The behaviour of solids within protoplanetary discs is key for understanding how planets form. Traditional models suggest that dust particles, initially sub-micron to micron in size, coagulate into aggregates that grow to 'pebble' sizes \citep[e.g.][]{Birnstiel_2024}. Growth can be halted by such barriers as the fragmentation barrier \citep{Dominik_1997, Dullemond_Dominik_2005,Blum_Wurm_2008}, the bouncing barrier \citep{Zsom_2010_bouncing_barrier, Dominik_2024}, and the radial-drift barrier \citep{Whipple_1972, Weidenschilling_1977a, Brauer_2008}. Depending on their material properties, some solids are more strongly coupled to sub-Keplerian gas discs than others. This coupling is the main driving force of planetesimal formation, producing the kilometre-sized precursors of protoplanets. These can form via resonant drag instabilities \citep{Squire_Hopkins_2018, Magnan_2024A}, with the streaming instability being the most widely studied in the literature \citep{Youdin_2005, Johansen_2007, Magnan_2024B}. From this point onwards, mutual collisions between planetesimals further grow the bodies until the efficient phase of pebble accretion sets in \citep{Ormel_Klahr_2010, Johansen_Lambrechts_2017, Ormel_2017_Emering_Paradigm_Pebble_Accretion}. Pebble accretion is extremely efficient mainly for two reasons. First, the marginal coupling of pebbles to the gas greatly enhances the effective accretion radius, as it causes them to spiral inwards towards the protoplanet rather than being scattered. Second, the previously mentioned drift barrier provides a nearly constant supply of pebble-sized particles drifting inwards from the outer disc.

The conventional way to calculate accretion rates is via a particle approach, in an unperturbed\footnote{Here, unperturbed means that the gaseous disc is not influenced by the gravitational potential of the (proto-)planet. This work also uses the term 'static' disc to refer to the same phenomenon: an unperturbed disc.} gaseous disc \citep[e.g.][]{Visser_2016}. Pebbles are modelled as moving through the gas disc under the influence of gravity (from the star and planet) and drag. This method has been able to determine the effects of a planet's eccentricity on growth \citep{Liu_Ormel_2018}, a planet's angular momentum response \citep{Visser_2020}, and even the growth timescale of a possible companion \citep{Konijn_2023}.

The motion of solids in protoplanetary discs can also be modelled hydrodynamically by treating dust\footnote{Throughout this work, we use the terms \textit{dust}, \textit{solids}, and \textit{pebbles} interchangeably. The terms refer to all solids in the gaseous disc, which we treat as a pressureless fluid.} as a pressureless fluid. Such an approach has been applied to both grid-based methods \citep[e.g.][]{Paardekooper_Mellema_2006, Mignone_2007_PLUTO, FARGO3D, Stone_2020_athena, Huang_Bai_2022, Lesur_2023_idefix} as well as smoothed-particle hydrodynamics (SPH) frameworks \citep[e.g.][]{Laibe_Price_2012, Price_Laibe_2015, Price_2018_Phantom}. In contrast to Lagrangian particle schemes, the grid-based fluid approach avoids sampling limitations and particle shot noise \citep{Genel_et_al_2013, Commercon_et_al_2023} and enables momentum exchange between gas and solids to be handled straightforwardly in the solver.

Early investigations of pebble accretion in the context of hydrodynamical gas models perturbed by a planet were conducted by \cite{Morbidelli_2012}, using the FARGO code \citep{FARGO3D_Hydrodynamic}. Building on this legacy, using the modern FARGO3D code \citep{FARGO3D}, \cite{Chrenko_2024} investigated the pebble flows in a gas disc to study the torque on embedded planets in the 2D pebble accretion regime. In their study, 2D multi-fluid simulations were first performed with gas and pebbles. However, the smoothing of the planetary gravitational potential in these simulations prevented pebble accretion from being resolved. To overcome this limitation, the authors introduced a hybrid approach in which the pebbles are evolved as Lagrangian superparticles in a steady-state gaseous background obtained from hydrodynamical simulations. In this setup gas does not evolve simultaneously with the pebbles, and the back-reaction of the pebbles on the gas is therefore neglected. FARGO3D does, however, allow gas and multiple dust fluids to evolve self-consistently as a coupled system \citep{FARGO3D_Multifluid}.

To date, most pebble accretion models have generally assumed a monodisperse particle population, corresponding to a single dust size. Recent work has demonstrated that adopting a polydisperse description can significantly affect the early stages of planet formation, for instance in the acoustic resonant drag instability \citep{Paardekooper_Aly_2025a} or, more specifically, the streaming instability \citep{Paardekooper_2020, Matthijsse_2024, Paardekooper_Aly_2025b}. Since solids of different sizes drift at different velocities \citep{Weidenschilling_1977a}, the accretion of material in later stages is also expected to differ substantially between the mono- and polydisperse cases.

Only a handful of studies have directly examined polydisperse pebble accretion \citep{Lyra_2023, Andama_2022}. \citet{Lyra_2023} developed an analytical framework for polydisperse pebble accretion, finding that the onset of the Bondi (lower-mass) regime occurs at lower core masses than in the monodisperse case, reducing the need for planetesimal collisions to reach the pebble accretion phase. Their results indicate a modest decrease of 3/7 in accretion efficiency in the Hill (higher-mass) regime. An earlier investigation by \citet{Andama_2022}, based on a viscous 1D disc model, found that polydisperse Hill growth yields a higher final core mass ($M_\mathrm{iso}$).

In this study, we developed a hydrodynamical, self-consistent, multi-fluid framework that models pebble accretion rather than Lagrangian particles in a static gas-disc. With this method, we investigated how disc evolution, solid-to-gas back-reaction, and a realistic polydisperse pebble population modify the accretion process. 

This paper is structured as follows. In Sect. \ref{sec:analytical_approach}, we show how a fluid model for monodisperse pebble accretion works in an unperturbed disc. In Sect. \ref{sec:numerical_approach}, we discuss the perturbed and polydisperse model setup as well as the assumptions made. We then validate the framework with earlier studies on pebble accretion. In Sect. \ref{sec:results}, we describe the results of our numerical simulations, first for an evolving the gas disc in a monodisperse setup, followed by an examination of the polydisperse case. In Sect. \ref{sec:discussion}, we discuss the outcome and results, and in Sect. \ref{sec:conclusions}, we summarise the implications.

\section{A fluid model for pebble accretion in an unperturbed gas disc}
\label{sec:analytical_approach}

\subsection{Governing equations: An accreting planet in the disc}
\label{subsec:equilibrium}

Due to mass and momentum conservation, the governing equations of a 2D, monodisperse dusty disc consist of
\begin{align}
	\label{eq:massconvgas} \partial_t\Sigma_\mathrm{g} + \nabla \cdot \left(\Sigma_\mathrm{g}\mathbf{v}_\mathrm{g}\right) &= 0,\\
	\label{eq:momentumconvgas} \partial_t\mathbf{v}_\mathrm{g} + \left(\mathbf{v}_\mathrm{g}\cdot\nabla\right)\mathbf{v}_\mathrm{g} &= -\nabla\Phi - \frac{\Sigma_\mathrm{d}}{\Sigma_\mathrm{g}}\frac{\mathbf{v}_\mathrm{g} - \mathbf{v}_\mathrm{d}}{\tau_\mathrm{s}} - \frac{\nabla P}{\Sigma_{\rm g}} + \frac{\nabla\cdot \mathsf{T}}{\Sigma_{\rm g}},\\
	\label{eq:massconvdust}\partial_t\Sigma_\mathrm{d} + \nabla \cdot \left(\Sigma_\mathrm{d}\mathbf{v}_\mathrm{d}\right) &= \nabla\cdot\left(D\Sigma_{\rm g}\nabla\left(\frac{\Sigma_{\rm d}}{\Sigma_{\rm g}}\right)\right),\\
    \label{eq:momentumconvdust}\partial_t\mathbf{v}_\mathrm{d} + \left(\mathbf{v}_\mathrm{d}\cdot\nabla\right)\mathbf{v}_\mathrm{d} &= -\nabla\Phi - \frac{\mathbf{v}_\mathrm{d} - \mathbf{v}_\mathrm{g}}{\tau_\mathrm{s}},
\end{align}
where $\Sigma$ denotes the surface density, $\mathbf{v}$ the velocity, $\tau_\mathrm{s}$ the stopping time, $\Phi=\Phi_\star+\Phi_\mathrm{pl}$ the gravitational potential (star and planet), $\nabla P$ the gas pressure gradient, and the subscripts g and d refer to gas and dust, respectively. We used a locally isothermal equation of state, $P=c_{\rm s}^2(r)\Sigma_{\rm g}$ and sound speed $c_{\rm s} \propto r^{-1/2}$ so that the disc has a constant aspect ratio $H_{\rm g}/r$, where $H$ is the scale height. The viscous stress tensor, $\mathsf{T}$, is given by
\begin{align}
    \mathsf{T} = \Sigma_{\rm g} \nu \left[\nabla {\bf v}_{\rm g} +\left(\nabla {\bf v}_{\rm g} \right)^T - \frac{2}{3}\mathsf{I} \nabla\cdot {\bf v}_{\rm g}\right],
\end{align}
where $\nu$ is the kinematic viscosity. The dust continuity equation includes dust diffusion with a diffusion coefficient $D$, which we set to $D=\nu$ throughout this work (i.e. we assume a Schmidt number of unity for simplicity\footnote{In FARGO3D's standard dust diffusion module \citep{Weber_2019}, a Schmidt number of unity is used.}).

The planet potential is given by
\begin{equation}
	\label{eq:planetpotential}\Phi_{\rm pl} = -\frac{GM_\mathrm{pl}}{\sqrt{r^2 + r_0^2 - 2rr_0\cos{\left(\varphi-\varphi_0\right)} + \lambda^2}},
\end{equation}
where $M_\mathrm{pl}$ is the mass of the planet, the subscript 0 denotes the planet's location, and $\lambda$ is a gravitational potential smoothing parameter. The reason this smoothing factor is necessary in a fluid approximation is two-fold. First, it eliminates the problem of a (close-to) infinite potential close by the gravitational source. Second, it provides an analogy for a 3D representation in a 2D simulation, i.e. when a parcel of gas is high in the disc yet close by in the $(r,\varphi)$-plane. However, the pebble layer is much thinner ($H_\mathrm{p}\ll H_\mathrm{g}$) and for most Stokes numbers considered in this work, remains well below the accretion radius for the planets in our parameter space ($M_\mathrm{pl}>1.5\ M_\oplus$).

For both problems, the smoothing factor creates a realistic solution; however, it misrepresents the mechanics close to the planet in a pebble accretion scenario (see Fig. 3 from \cite{Chrenko_2024} for reference). Thus, for pebble accretion we want this factor $\lambda$ to be as small as possible. 

Pebble accretion is such a highly efficient phase of growth because the accretion radius $R_\mathrm{acc}$ greatly exceeds planetary radius. In the Hill regime, which is the focus of this work, $R_\mathrm{acc}$ \citep{Ormel_2017_Emering_Paradigm_Pebble_Accretion} can be defined as
\begin{equation}
\label{eq:R_acc} R_\mathrm{acc} = \left(\frac{GM_\mathrm{pl}\tau_\mathrm{s}}{\Omega}\right)^{1/3} \approx R_\mathrm{Hill}\mathrm{St}^{1/3},
\end{equation}
where $\Omega$ is the angular velocity, $R_\mathrm{Hill}=a_\mathrm{pl} \left(M_\mathrm{pl}/3M_\star\right)^{1/3}$ is the Hill radius \citep{Hill_1878}, and $a_\mathrm{pl}$ is the orbital distance of the planet.

\subsection{Dust streamlines}
\label{subsec:planet_pertubation}

\begin{figure}[]
   \centering
   \includegraphics[width=\columnwidth]{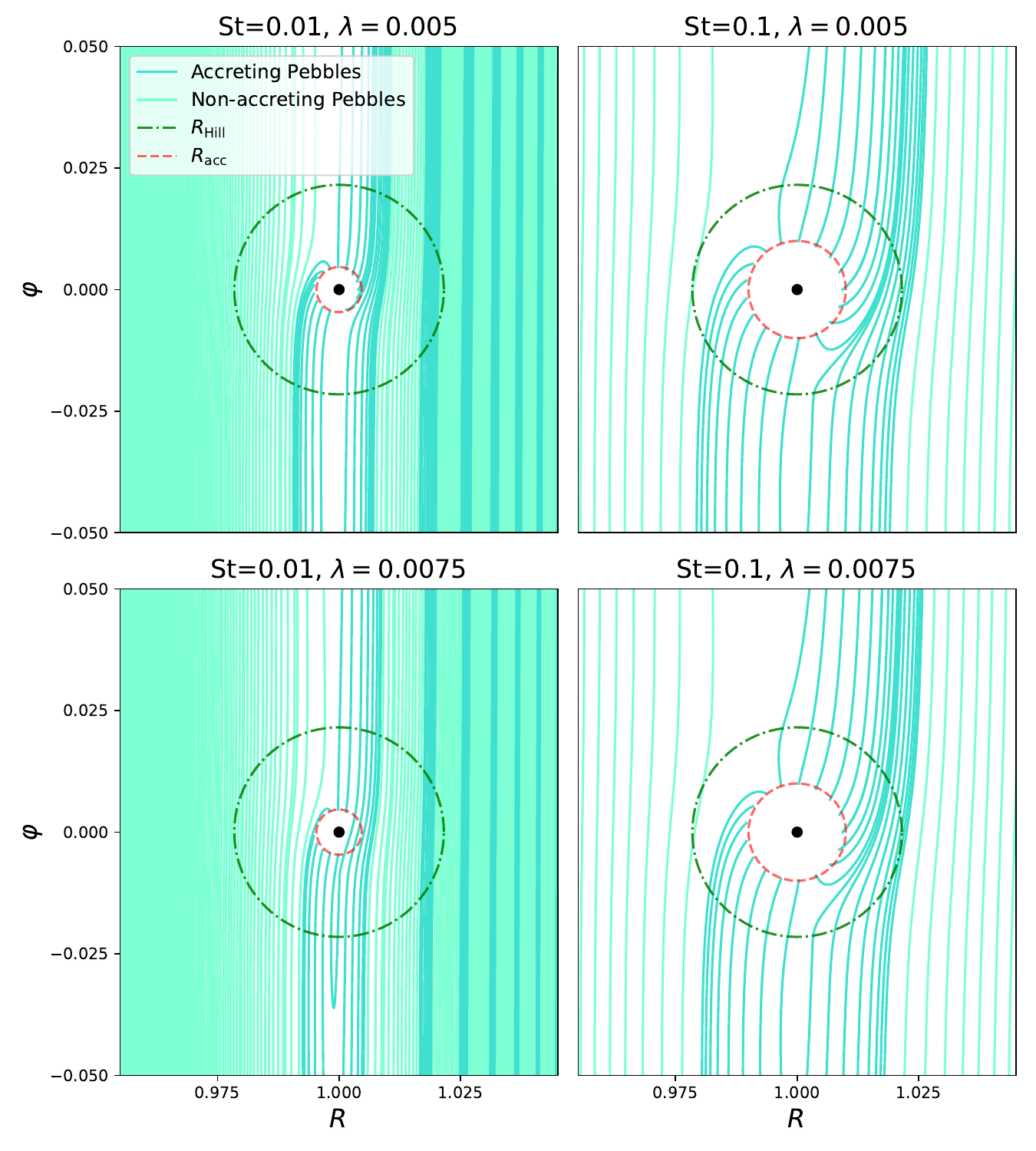}
   \caption{Velocity fields for pebbles in an unperturbed gaseous disc, calculated by Fourier-decomposing the planet's potential (as explained in Appendix \ref{apx:Velocityfield}). Here, a 10 $M_\oplus$ planet is embedded at 1 AU around a 1 $M_\odot$ star. Two Stokes numbers, $\mathrm{St}=0.01$ (left) and $\mathrm{St}=0.1$ (right), as well as two different softening parameters, $\lambda=0.005$ (top) and $\lambda=0.0075$ (bottom), are shown. The darker shade indicates the accreted pebbles, the dash-dotted green circle indicates the Hill radius, and the dashed red circle indicates the accretion radius $R_\mathrm{acc}$. Only one in five trajectories of the non-accreting pebbles is plotted to avoid overfilling the figure with streamlines.}
   \label{fig:vel_fields}
\end{figure}

In this section, we assumed that the gas disc is static, inviscid, and does not feel the planet -- a common assumption in pebble accretion studies \citep[e.g.][]{Ormel_Klahr_2010, Lambrechts_Johansen_2012, Visser_2016}. An axisymmetric steady solution can be constructed where the gas velocity $\mathbf{v}_\mathrm{g}$ is purely azimuthal. The equilibrium gas velocity can be found by balancing gravity and the radial pressure gradient. This enables a  description of the azimuthal gas velocity by relating it to the Keplerian velocity $v_\mathrm{K}$ via a non-dimensional value, $\eta$:
\begin{equation}
	\label{eq:vg_vk} v_{\mathrm{g},\varphi} = v_\mathrm{K}\left(1-\eta\right),
\end{equation}
with
\begin{equation}
	\label{eq:eta} \eta = -\frac{1}{2}\frac{c_\mathrm{s}^2}{v_\mathrm{K}^2}\frac{\mathrm{d} \ln{P}}{\mathrm{d} \ln{r}}\approx \left(\frac{H_\mathrm{g}}{r}\right)^2.
\end{equation}
Note that $\eta\ll1$, so that $v_{{\rm g},\varphi} \approx v_{\rm K}$. 

Unlike the gas, the dust does feel the planet's influence, but we assume this effect to be small. The dust momentum equation (Eq. \ref{eq:momentumconvdust}) is given by
\begin{align}
\partial_t\mathbf{v}_\mathrm{d} + \left(\mathbf{v}_\mathrm{d}\cdot\nabla\right)\mathbf{v}_\mathrm{d} &= -\nabla\Phi_*  - \frac{\mathbf{v}_\mathrm{d} - {\bf v}_{\rm K} }{\tau_\mathrm{s}}+ \frac{\mathbf{v}_\mathrm{g} - {\bf v}_{\rm K}}{\tau_\mathrm{s}}-\nabla\Phi_{\rm pl},
\end{align}
where the last two terms are treated as small perturbations. As with the gas, we seek a steady state. Without the perturbations, the dust equilibrium velocity is purely azimuthal, with $v_{{\rm d},\varphi}=v_{\rm K}$. 

For simplicity, we assumed ${\rm St} \equiv \Omega\tau_{\rm s}$, the Stokes number, to be constant. The perturbed dust velocity field can be found by linearising the dust momentum equation, Fourier-decomposing the planet potential, and adding all contributions (see Appendix \ref{apx:Velocityfield}). The number of Fourier modes necessary for accurate results depends on the smoothing length $\lambda$; smaller values of $\lambda$ require more modes. From the resulting velocity field, dust streamlines are calculated. 

In absence of the planet, we find that the perturbed velocity field is 
\begin{align}
	\label{eq:velPert0} v_r' &= \frac{2r\mathrm{St}\left(\Omega_\mathrm{g} - \Omega_\mathrm{K}\right)}{1 + \mathrm{St}^2},\\
	\label{eq:freqPert0} \Omega' &= \frac{\Omega_\mathrm{g} - \Omega_\mathrm{K}}{1 + \mathrm{St}^2},
\end{align}
which is the usual radial drift solution in an unperturbed gas disc \citep{Weidenschilling_1977a}. After including the planet's perturbation (Appendix \ref{apx:Velocityfield}), we integrated to find the velocity field of the pebbles. We show four different velocity fields in Fig. \ref{fig:vel_fields} for two different Stokes numbers (left to right) and two different softening parameters $\lambda$ (top to bottom). The difference in velocity fields for the different Stokes numbers (St) is easily noticeable: a higher St drifts faster while a lower St is more coupled to the gas. The latter also reduces the planet's influence. The difference for the smoothing factor $\lambda$ is much less visible; however, there is certainly a difference. We see fewer streamlines accreting with higher $\lambda$, which is not unexpected since it essentially lowers the gravitational potential (Eq. \ref{eq:planetpotential}). This effect is more pronounced for the lower St since it is more coupled to the gas.

\subsection{Measuring accretion efficiency: The impact of softening parameter}
\label{subsec:softening}

\begin{figure}[]
   \centering
   \includegraphics[width=\columnwidth]{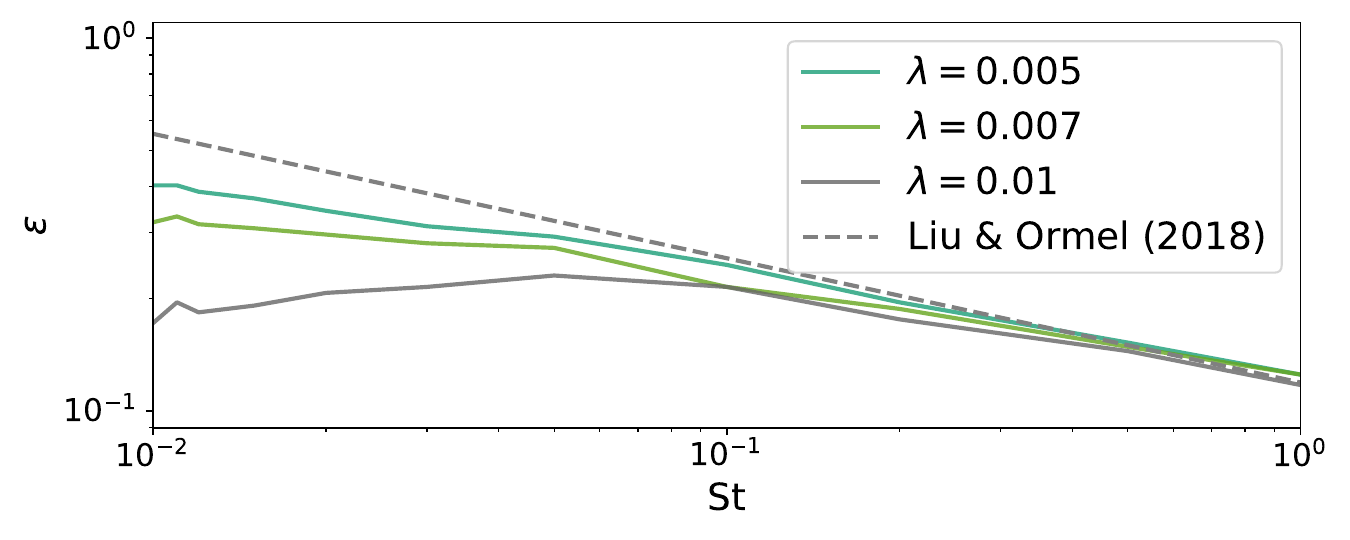}
   \caption{Accretion efficiency, $\varepsilon$, for a 10 $M_\oplus$ planet, as calculated by \cite{Liu_Ormel_2018} (dashed line), compared to the efficiency obtained with our method for different smoothing parameters, $\lambda$ (solid-coloured lines).}
   \label{fig:softening}
\end{figure}

One method for testing pebble accretion involves examining the accretion efficiency $\varepsilon$ \citep{Guillot_2014, Lambrechts_Johansen_2014, Liu_Ormel_2018}. This efficiency is the fraction of all radially drifting mass accreted by the planet:
\begin{equation}
	\label{eq:epsliondefinition} \varepsilon = \frac{\dot{M}_\mathrm{acc}}{\dot{M}_\mathrm{rad,peb}},
\end{equation}
where $\dot{M}_\mathrm{acc}$ is the pebble accretion rate and $\dot{M}_\mathrm{rad,peb}$ is the total radial mass flow of pebbles in the disc. We calculated this efficiency simply by examining the streamlines (Fig. \ref{fig:vel_fields}), specifically those that fall into the accretion radius $R_\mathrm{acc}$ of the planet (Eq. \ref{eq:R_acc}) and those that do not. We did this for different smoothing parameters $\lambda$ and for multiple St. We compare our findings to those of \cite{Liu_Ormel_2018} in Fig. \ref{fig:softening}.

Our findings are mostly consistent with the expectations of \cite{Liu_Ormel_2018}. However, once $\lambda$ increases, it begins to deviate at the lower end of the Stokes numbers, lowering the efficiency $\varepsilon$. Intuitively, this makes sense since for lower St the decisive 'accreting moment' occurs closer to the planet; therefore, it is more heavily impacted by the smoothing factor. This is also visible in Fig. \ref{fig:vel_fields}, where, for the lower St, fewer streamlines accrete onto the planet at $\lambda=0.0075$ than at $\lambda=0.005$.

\section{Numerical hydrodynamical model for pebble accretion}
\label{sec:numerical_approach}

We simulated the gas and dust in a global disc using the multi-fluid hydrodynamical FARGO3D code \citep{FARGO3D, FARGO3D_Multifluid}. This is a finite-difference, grid-based hydrodynamical code capable of simulating multiple dust species with an evolving gas disc in multiple geometries. Because we are working in a disc, we used cylindrical coordinates $(r,\varphi,z)$. Since we primarily examine a phase where 2D accretion takes place, i.e. the accretion radius surpasses the pebble scale height, we used a vertically integrated 2D setup in the $(r,\varphi)$ grid. Opting for a 2D rather than 3D simulation directly helps keep the computation time manageable.

For the duration of this work, we focus on a 1 $M_\odot$ star, with a gas surface density at the location of the planet of $\Sigma_\mathrm{g}\ (a_\mathrm{pl}=1\ \mathrm{AU})=10^{-4}\ M_\oplus\ \mathrm{AU}^{-2}$. The disc viscously evolves with an $\alpha$-description of $\alpha=10^{-3}$. We obtain a constant aspect ratio of $H_\mathrm{g}/r=0.05$. The inner and outer radii of our simulation are situated at 0.6 AU and 1.6 AU, respectively. Our grid has dimensions of \texttt{400x2512} in the radial and azimuthal direction, respectively, such that the grid cells around the planet are square. Note that the code essentially uses dimensionless code units, but we adopt $M_\odot$, AU, and $\left(GM_\odot/\mathrm{AU}^3\right)^{-1/2}$ as the mass, distance, and time units in all our figures. All our results are therefore scalable with ease, we only keep the aforementioned units to make our figures more easily comprehensible for the reader.

We did not adopt a prescribed or constant pebble mass flux at the radial boundaries. Instead, at both the inner and outer radial boundaries, we applied standard Keplerian disc boundary conditions. In these, the density and azimuthal velocity of each fluid are relaxed to the analytic background disc profiles, while the radial velocity is antisymmetric, corresponding to a no-penetrating condition. These boundaries therefore act as a reservoir that maintains the background disc state near the domain edges without enforcing a fixed inflow rate. The pebble mass flux is thus not externally controlled but emerges self-consistently from the interaction between the disc and the planet. The reasoning behind this method is as follows: Imposing a constant pebble mass flux in a polydisperse disc is non-trivial, since different grain sizes experience distinct drift velocities and filtering efficiencies; therefore, we adopt these reservoir-type boundary conditions to allow the mass flux of each species to adjust self-consistently.

\subsection{Accretion mechanism}
\label{subsec:acc_mechanism}

Previous studies have raised concerns about whether pebble accretion can be accurately reproduced within a fluid approximation. \cite{Chrenko_2024} for instance found multi-fluid simulations to be impractical, mainly due to the smoothing of the planetary gravitational potential. These difficulties are demonstrated in their Fig. 3.

In this work we adopted an alternative strategy: the gas retains its usual gravitational smoothing ($\lambda_\mathrm{g} = 0.6 H_{\rm g}$) to account for its vertical scale height, but each dust fluid experiences an unsmoothed potential ($\lambda_\mathrm{d} = 0$). This is valid because the pebble scale height $H_\mathrm{p}$ is mostly smaller than the accretion radius in our parameter space \citep{Birnstiel_2024, Fromang_Nelson_2009, Binkert_2023}, implying 2D accretion. We can see this by assuming the scale height of the pebbles to be
\begin{equation}
\label{eq:scaleheight_pebbles} H_\mathrm{p} = H_\mathrm{g}\sqrt{\frac{\alpha}{\mathrm{St}+\alpha}},
\end{equation}
where the maximum scale height corresponds to our smallest Stokes number grain, i.e. $\mathrm{St}_\mathrm{min}=10^{-2}$, in our case. This would correspond to a pebble scale height, which remains smaller than the Hill radius for our lowest simulated planet mass  of 1.5 $M_\oplus$.

This is consistent with a very thin pebble disc, which does not necessitate smoothing to account for vertical height. For every dust fluid, we created a separate potential field\footnote{The reason we have a separate potential field for all dust fluids is because we let $\Phi_\mathrm{pl}$ be constant for all cells within $R_\mathrm{acc} - l_\mathrm{cell}$, where $l_\mathrm{cell}$ is the length of one grid cell. This avoids the (near-) infinite potential and keeps the potential gradient as low as possible, which in turn makes the variable time-step manageable. Since $R_\mathrm{acc}$ depends on St, we had to redefine it for every single dust species.} without $\lambda$. After which, we removed all the solid mass within $R_\mathrm{acc}$. 

Previous attempts to replicate pebble accretion in fluid simulations \citep{Chrenko_2017,Regaly_2020,Pierens_2023} were based either on gas-accretion schemes analogous to those of \cite{Kley_1999} or on semi-analytical accretion prescriptions derived from particle-based studies \citep{Liu_Ormel_2018,Ormel_Liu_2018}. In contrast, our approach is more targeted towards pebble accretion by removing all material within $R_\mathrm{acc}$, which we argue more accurately captures the essence of pebble accretion since all mass within the accretion radius is captured in that phase by the planet. Since we modelled the pebbles as a pressureless fluid, the streamlines outside $R_\mathrm{acc}$ remain unaffected by the accretion process.

\subsection{Validating the approach: Comparison to earlier studies}
\label{subsec:comparison_PA_studies}   

\begin{figure}[]
   \centering
   \includegraphics[width=0.495\columnwidth]{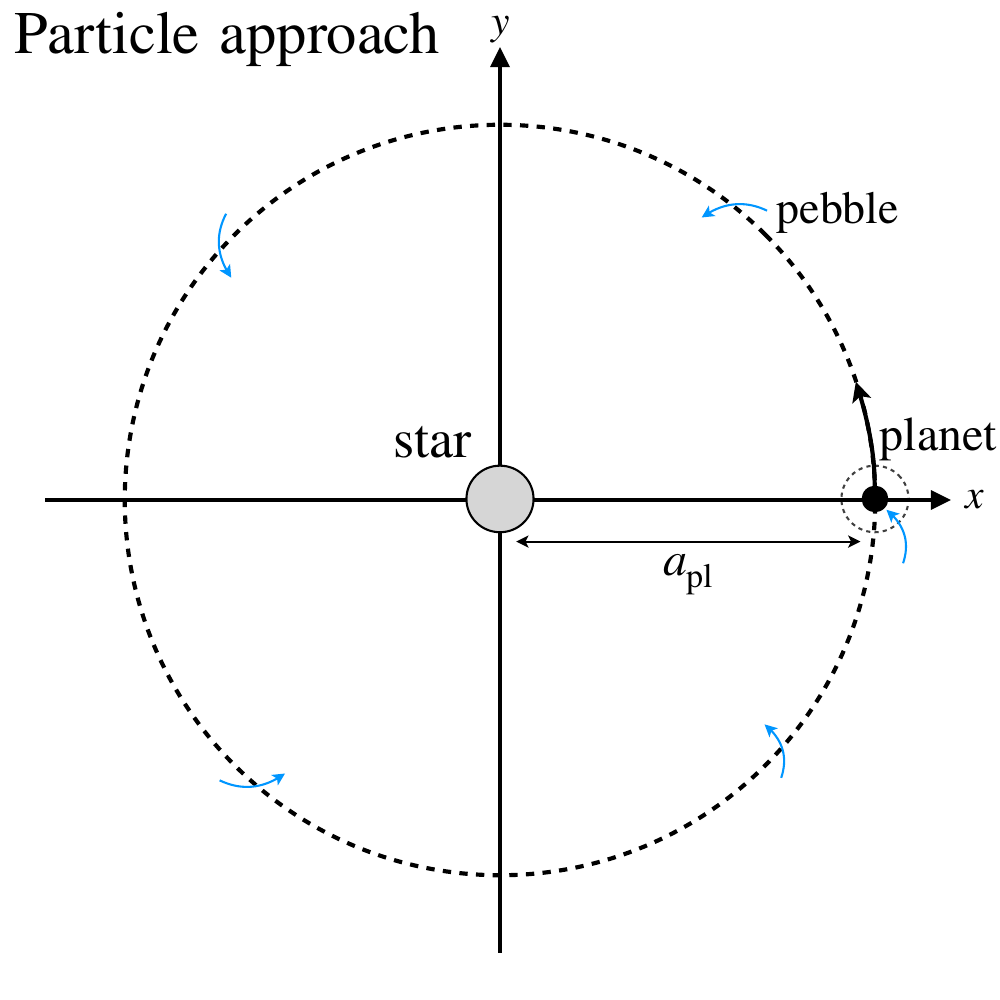}
   \includegraphics[width=0.495\columnwidth]{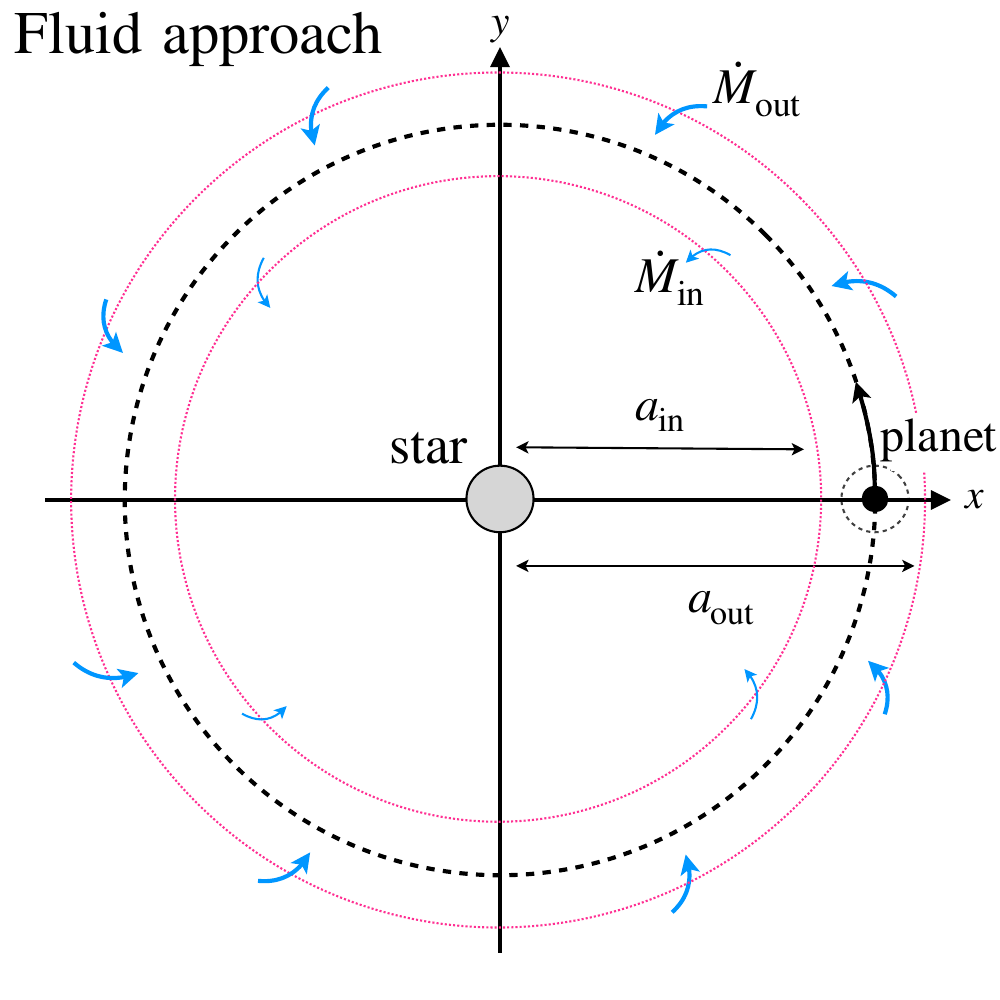}
   \caption{\textit{Left:} Sketch of the accretion efficiency determined by \cite{Liu_Ormel_2018}. Individual pebbles simulated either hit the planet and accrete or pass by the ring at $r=a_\mathrm{pl}$. The efficiency is calculated using Eq. \ref{eq:acc_eff_epsilon_LO18}. \textit{Right:} Sketch of the accretion efficiency, as determined by our method. The dust radially drifts inwards with a mass flow $\dot{M}$, calculated at two rings located at $a_\mathrm{in}$ and $a_\mathrm{out}$ using Eq. \ref{eq:mdot_disc}. The efficiency $\varepsilon$ is subsequently calculated using Eq. \ref{eq:acc_eff_epsilon}.}
   \label{fig:sketch_epsilon}
\end{figure}

To evaluate our pebble accretion mechanism, we benchmarked it against earlier studies. \cite{Liu_Ormel_2018} provides a useful point of comparison. They evaluated the accretion efficiency, $\varepsilon$ (Eq. \ref{eq:epsliondefinition}), both locally and globally. Their work employed the conventional particle-based approach in a unperturbed gas disc. They defined the accretion efficiency as the ratio of the number of pebbles accreted over the total number of pebbles entering from the outer disc: 
\begin{equation}
	\label{eq:acc_eff_epsilon_LO18} \varepsilon_\mathrm{LO18} = \frac{N_\mathrm{hit}}{N_\mathrm{total}},
\end{equation}
where $N_\mathrm{hit}$ is the amount of pebbles accreted and $N_\mathrm{total}$ is the total amount of pebbles reaching the planet from the outer ring. A schematic of this method is shown on the left of Fig. \ref{fig:sketch_epsilon}.

Since we employed fluid approximation, we required a different method to calculate the accretion efficiency. Due to our chosen disc parameters, the radial mass flow $\dot{M}$ should be constant in $r$ if no planet is embedded. When an accreting planet is added, the mass removal ensures $\left|\dot{M}_\mathrm{in}\right|<\left|\dot{M}_\mathrm{out}\right|$, where $\dot{M}_\mathrm{in}$ and $\dot{M}_\mathrm{out}$ are the total radial mass flows through rings situated in and outside the planet, respectively. A sketch of this method is shown on the right of Fig. \ref{fig:sketch_epsilon}. We easily calculate the efficiency from these values via
\begin{align}
	\label{eq:acc_eff_epsilon} \varepsilon &= \frac{\dot{M}_\mathrm{out} - \dot{M}_\mathrm{in}}{\dot{M}_\mathrm{out}},
\end{align}
where the pebble radial mass flow in a disc is
\begin{equation}
	\label{eq:mdot_disc} \dot{M}_\mathrm{rad}(r) = \int_\mathrm{ring} d\dot{M}_\mathrm{rad} = \int_0^{2\pi}\Sigma(r,\varphi) v_r(r,\varphi) r d\varphi
\end{equation}
and $v_r$ is the radial component of the dust velocity $\mathbf{v}_\mathrm{dust}$. We note that Eq. \ref{eq:mdot_disc} describes the advective (bulk) radial mass flux for pebbles. In addition to this advective component, a diffusive flux arises due to the diffusion term in Eq. \ref{eq:massconvdust}. In the present work, we do not explicitly analyse the diffusive contribution. In the absence of sharp gradients, the ratio of diffusive to advective flux is $\sim\alpha/\mathrm{St}$, meaning that for our parameters, the advective flux is always more than an order of magnitude larger. Once we reach the pebble isolation mass, small grains partially diffuse across the pressure bump generated by the planet \citep[e.g.][]{Bitsch_2018, Ataiee_2018}, thereby modifying the net flux. In this work, however, we do not analyse this effect.

The orbital distances of $\dot{M}_\mathrm{in}$ and $\dot{M}_\mathrm{out}$ are located at $a_\mathrm{in}$, and $a_\mathrm{out}$, respectively. We chose $a_\mathrm{in}$ and $a_\mathrm{out}$ carefully to truly capture the difference between the incoming and passing mass flow. Choosing values too close to the planet for instance could artificially increase or decrease both the mass flow and the efficiency. Doing so via this method, we obtain the same results if we measure efficiency by comparing the removed disc mass with the incoming mass flow. 

\begin{figure}[]
   \centering
   \includegraphics[width=\columnwidth]{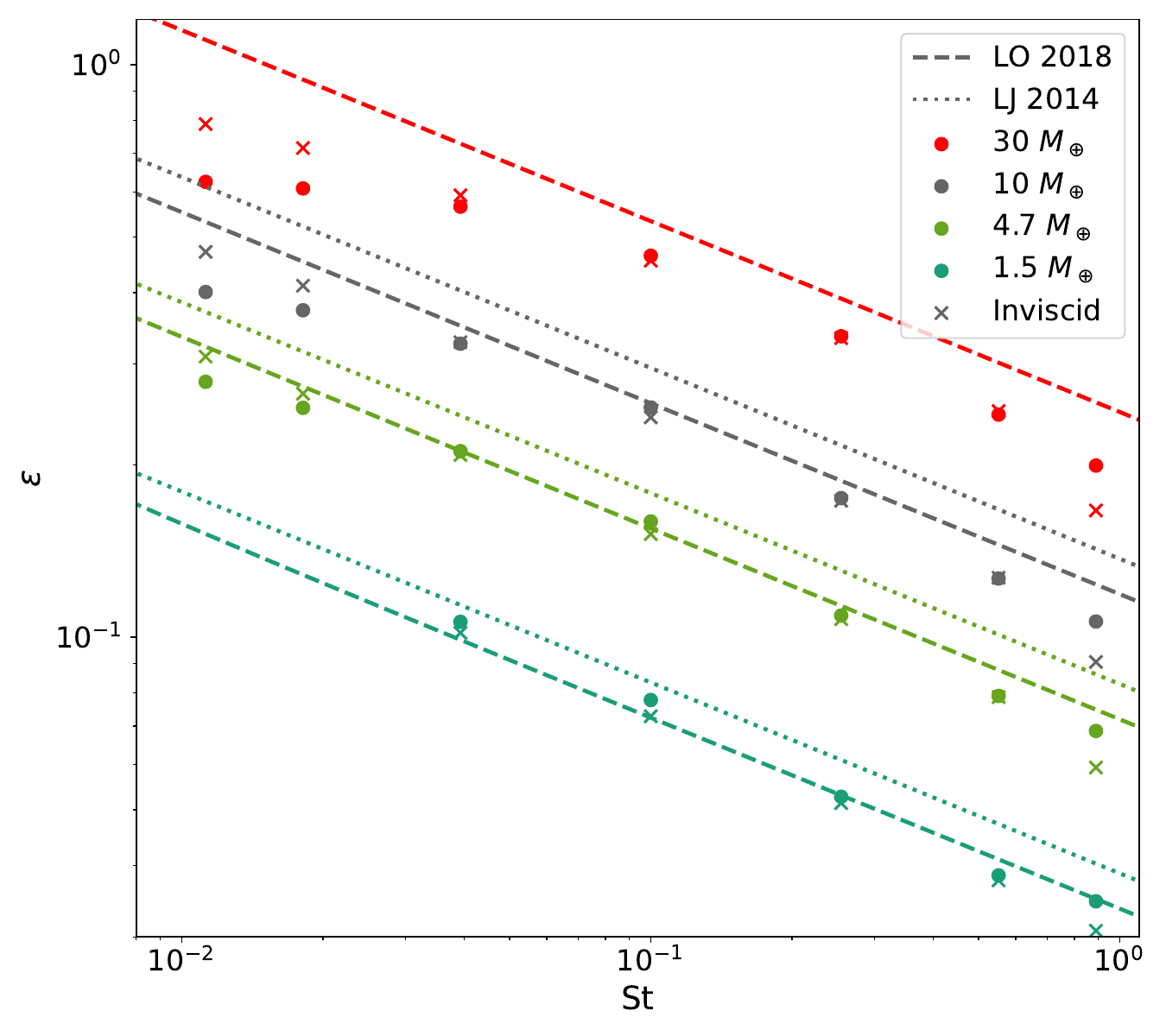}
   \caption{Accretion efficiency, $\varepsilon$, for multiple planetary masses in a disc where the gas is unperturbed by the planet and back-reaction is disabled. The results are taken from the repo package of \cite{Liu_Ormel_2018} (dashed lines); Eq. 33 of \cite{Lambrechts_Johansen_2014} (dotted lines); our simulation for a viscous disc with $\alpha=10^{-3}$ (dots); and an inviscid disc (crosses). Note that the smallest two Stokes numbers are not plotted for the 1.51 $M_\oplus$ planet because the accretion radius for these Stokes numbers is smaller than the cell width at our resolution.}
   \label{fig:epsilons_monodisperse}
\end{figure}

We compare our simulation to \cite{Liu_Ormel_2018}'s (dashed lines) in Fig. \ref{fig:epsilons_monodisperse}. For completeness we also show Eq. 33 of \cite{Lambrechts_Johansen_2014} (dotted lines). We use an unperturbed gas disc, consistent with the classical particle-based pebble accretion setup. We use a gas pressure gradient of $\eta=0.001875$. The gas is kept unperturbed by allowing only the dust fluids feel the planetary potential (i.e. $\Phi_\mathrm{pl}=0$ for the gas), so the gas remains static while the dust responds to the planet. Back-reaction is disabled to prevent the dust from influencing the gas. 

We simulated four different planet masses -- 1.5, 4.7, 10, and 30 $M_\oplus$ -- for completeness. Note that the latter planetary mass achieves an efficiency of $\varepsilon>1$ in line with the expectation of \cite{Liu_Ormel_2018}. We therefore included it mostly as a sanity check. The two studies included here for comparison use an inviscid disc, which is why we also account for inviscid simulations in our comparison. It is clear that the inviscid disc more closely resembles the expected values from \cite{Liu_Ormel_2018}. Our viscous simulation shows a slight discrepancy for lower Stokes number pebble, which is not unexpected. For ${\rm St} \lesssim H_{\rm g}^2/r^2$, the viscous radial gas speed exceeds the pebble drift speed, likely affecting pebble accretion. We note lower efficiencies for the higher Stokes numbers compared to \cite{Liu_Ormel_2018}. We attribute this to the fact that the power law relies on the low-Stokes approximation (their Appendix A.1), which our simulations do not use.

This comparison demonstrates the promising results of our method for intermediate planetary masses ($M_\mathrm{pl}\ \tilde{\in}\ [1,10]\ M_\oplus$), as well as its necessity for higher masses ($M_\mathrm{pl}\gtrsim10M_\oplus$). The latter is because the local particle approach can overestimate the efficiency (exceeding 1), and neither the local nor the global particle approach account for the perturbations of the planet on the gas disc (which we address in Sect. \ref{subsec:Mono_perturbing}). We note, however, the shortcomings of our method for lower planetary masses; even for 1.5 $M_\oplus$, the smallest two Stokes number grain sizes are not accreted since $R_\mathrm{acc}$ is smaller than the grid-cell size of our simulation. At these masses the gas is insignificantly perturbed (see Sect. \ref{subsec:Mono_perturbing}), so the particle-based method is expected to be sufficient.

\subsection{Approximating the size distribution for a polydisperse disc}
\label{subsec:GL_Conjecture}

\begin{figure}[]
   \centering
   \includegraphics[width=\columnwidth]{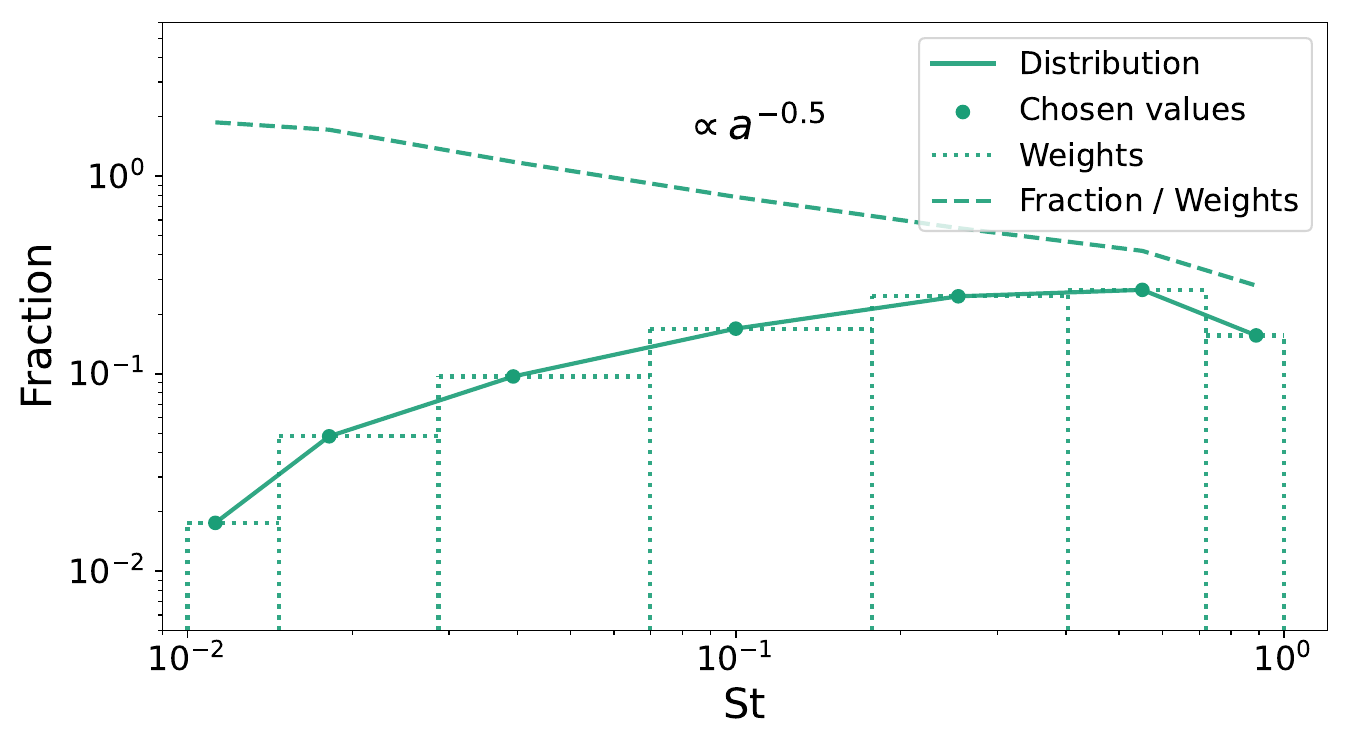}
   \caption{Distribution of density weights over different St numbers for an MRN-distribution spanning $\mathrm{St}\in\left[10^{-2},10^0\right]$. The dots represent the calculated St values, and the dotted lines showcase the weights that every dot represents. The dashed line shows the density fraction divided by the weight of the Stokes number, closely following the MRN-distribution ($\propto a^{-0.5}\propto\mathrm{St}^{-0.5}$).}
   \label{fig:St_Distribution}
\end{figure}

In our polydisperse simulations, we considered a continuous size distribution. This is computationally challenging with a finite number of species. We tackled this problem by means of the Gauss-Legendre (GL) quadrature. 

Following the methods of \cite{Paardekooper_2020}, we first defined a 'size density' $\sigma$, such that
\begin{equation}
	\label{eq:size_density} \Sigma_\mathrm{d} = \int \sigma(\mathrm{St})\mathrm{dSt},
\end{equation}
where $\Sigma_\mathrm{d}$ is the total surface density of the solids. The problem lies in the drag term in Eq. \ref{eq:momentumconvgas} (second term on the right-hand-side). For a polydisperse setup this becomes an integral, which we approximated as
\begin{equation}
	\label{eq:GL_integral} \frac{1}{\Sigma_\mathrm{g}}\int\sigma\frac{\mathbf{v}_\mathrm{d}-\mathbf{v}_\mathrm{g}}{\mathrm{St}}\mathrm{dSt} \approx \frac{1}{\Sigma_\mathrm{g}}\sum_n w_n\sigma(\mathrm{St}_n)\frac{\mathbf{v}_\mathrm{d}(\mathrm{St}_n)-\mathbf{v}_\mathrm{g}}{\mathrm{St}_n},
\end{equation}
with weights $w_n$ and nodes $\mathrm{St}_n$. These integration nodes and weights ($\mathrm{St}_n,w_n$) correspond to the roots and coefficients of the Legendre polynomial of order $n$. This approach minimises integration errors compared to simple midpoint or logarithmic binning schemes. \cite{Matthijsse_2024} showed a lognormal distribution approximated with an error of $\sim10^{-4}$ using only five bins with the GL-method (their Fig. 1). By contrast, this same amount of bins gives an error of $\sim10^{-2}$ using a log-uniform method. Furthermore, the growth rate they measured shows higher accuracy for a GL method with five bins, compared to the log-uniform distribution with 40 bins (their Table 2).

Following \cite{Matthijsse_2024}, we generated the nodes and weights using the \texttt{scipy.special.roots.legendre} routine \citep{Scipy} and applied them to an MRN-like dust size distribution \citep{Mathis_1977, Draine_1984} with $F_\mathrm{MRN}(a)\propto a^{-3.5}$, corresponding to a mass distribution of $\propto a^{-0.5}\propto\mathrm{St}^{-0.5}$. We sampled Stokes numbers in the range $\mathrm{St}\in\left[10^{-2},10^0\right]$. This distribution is shown in Fig. \ref{fig:St_Distribution}. Note that, due to the logarithmic $x$-axis and the chosen weights, it might not appear to be an MRN-distribution, which is why we show the dashed line as a visual aid.

If we then define the densities of the individual pebble species as
\begin{equation}
	\Sigma_{\mathrm{d},n}\equiv w_n\sigma(\mathrm{St}_n),
\end{equation} 
we can use the back-reaction sum in FARGO3D and thereby simulate an analogue to a continuous distribution, albeit technically discrete.

\section{Results}
\label{sec:results}

We ran our simulation for different scenarios to compare the full effect of an evolving gaseous disc, dust-to-gas feedback, and a polydisperse disc. First, we ran our original monodisperse simulations with and without an evolving gaseous disc (Sect. \ref{subsec:Mono_perturbing}) and dust-to-gas feedback (Sect. \ref{subsec:Mono_feedback}). We then introduced a fiducial model for polydisperse accretion with an evolving gas disc in the Hill regime (Sect. \ref{subsec:Polydisperse}) and compared our findings with previous analytical (polydisperse) results (Sect. \ref{subsec:comparison_analyticalPPA}).

\subsection{Monodisperse: Perturbing the gas disc}
\label{subsec:Mono_perturbing}

\begin{figure}
    \centering
    \includegraphics[width=\columnwidth]{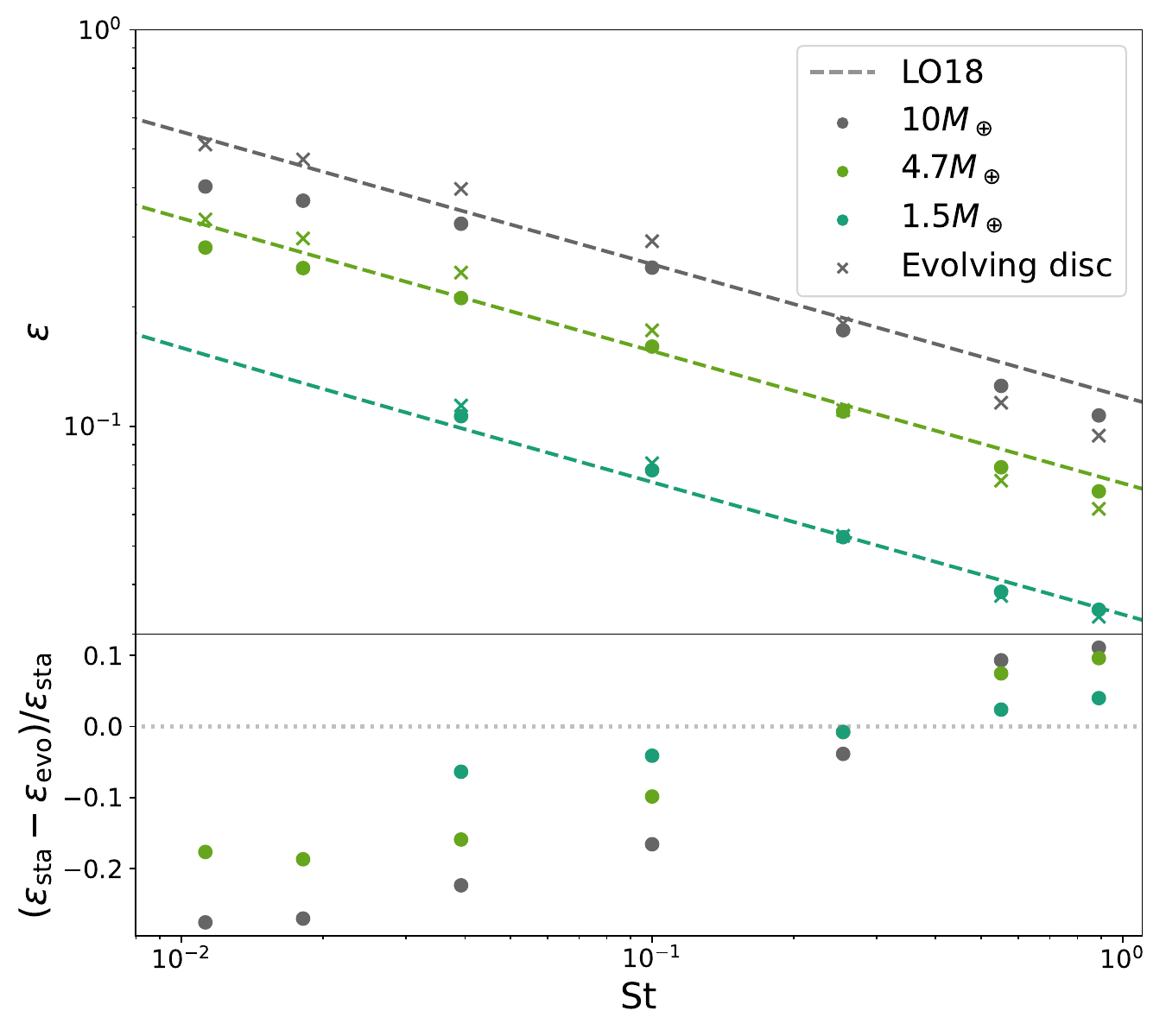}
    \caption{\textit{Top:} Accretion efficiency, $\varepsilon$, for three different planet masses for a static (dots) and evolving disc (crosses). Overplotted (dashed lines) are the results of \cite{Liu_Ormel_2018}, which represent static discs. \textit{Bottom:} Relative difference between $\varepsilon_\mathrm{sta}$, and $\varepsilon_\mathrm{evo}$, the accretion efficiency for a static and evolving disc, respectively. Note that the evolving disc is more efficient for the lower Stokes numbers.}
    \label{fig:efficiency_sta_evo}
\end{figure}

For the classic case of monodisperse pebble accretion, we turned on the evolving gas disc for the same (viscous) simulations as in Fig. \ref{fig:epsilons_monodisperse} to examine its effect on the accretion efficiency. Our result are shown in Fig. \ref{fig:efficiency_sta_evo}.

\begin{figure*}
    \centering
    \includegraphics[width=0.497\textwidth]{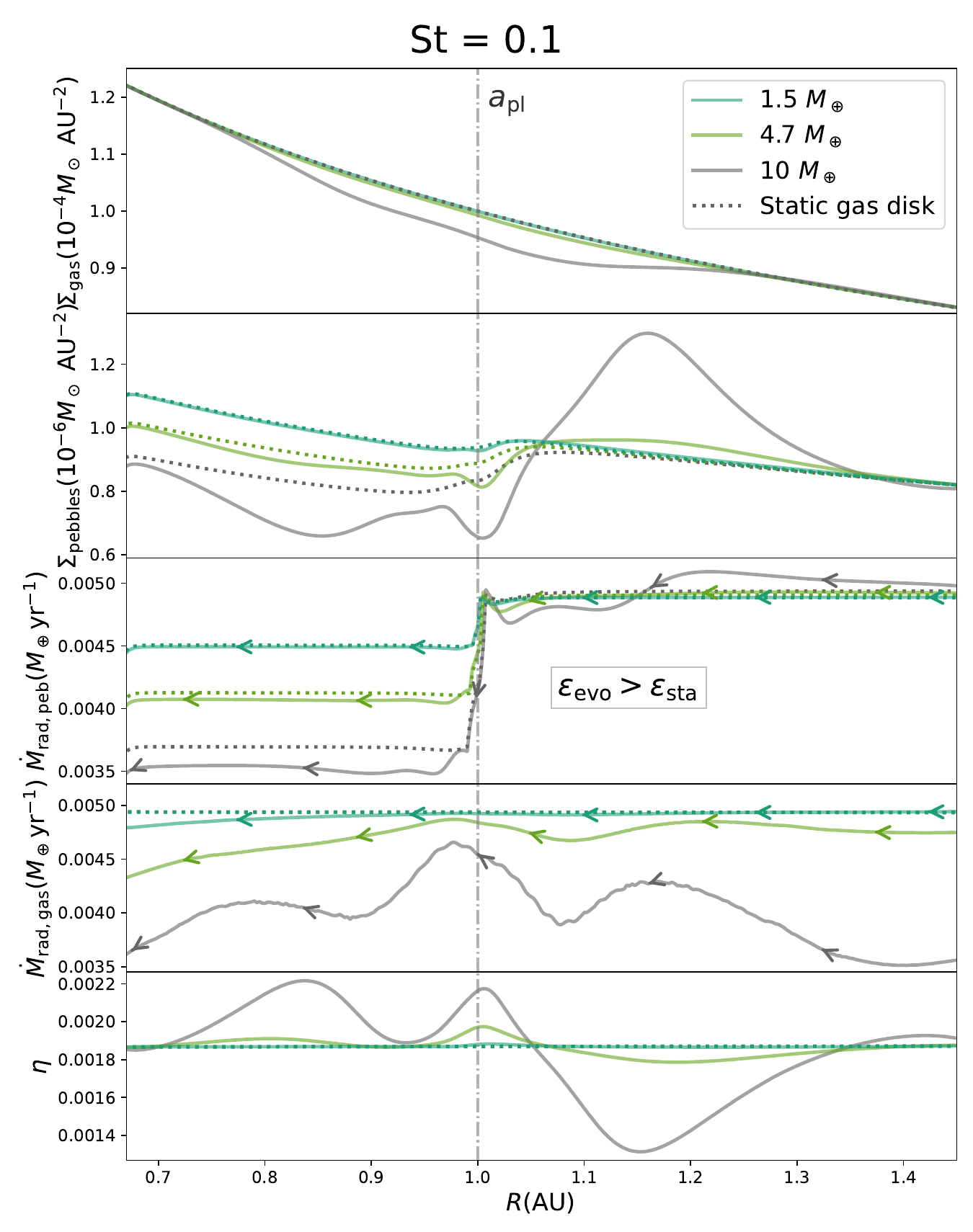}
    \includegraphics[width=0.497\textwidth]{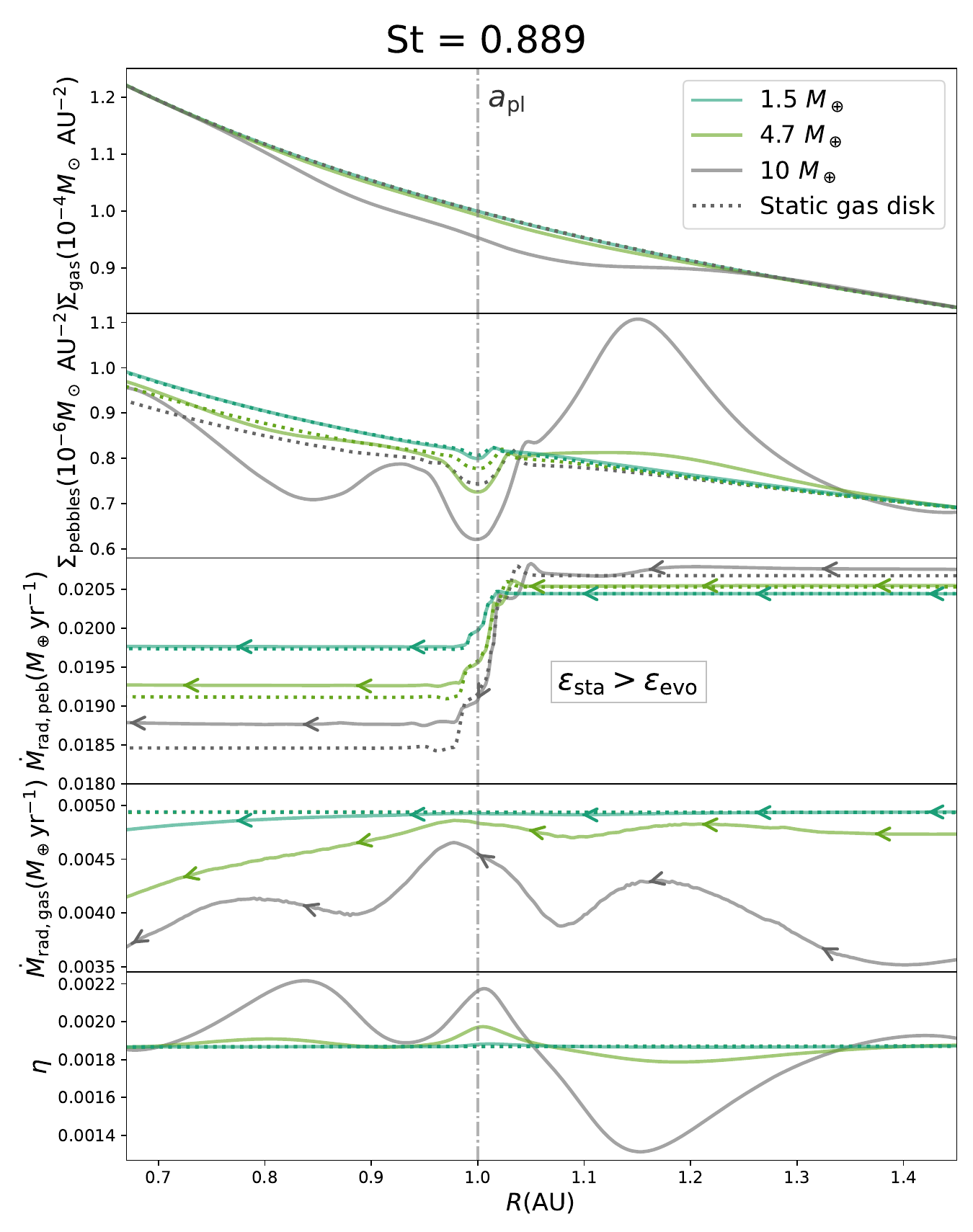}
    \includegraphics[width=0.497\textwidth]{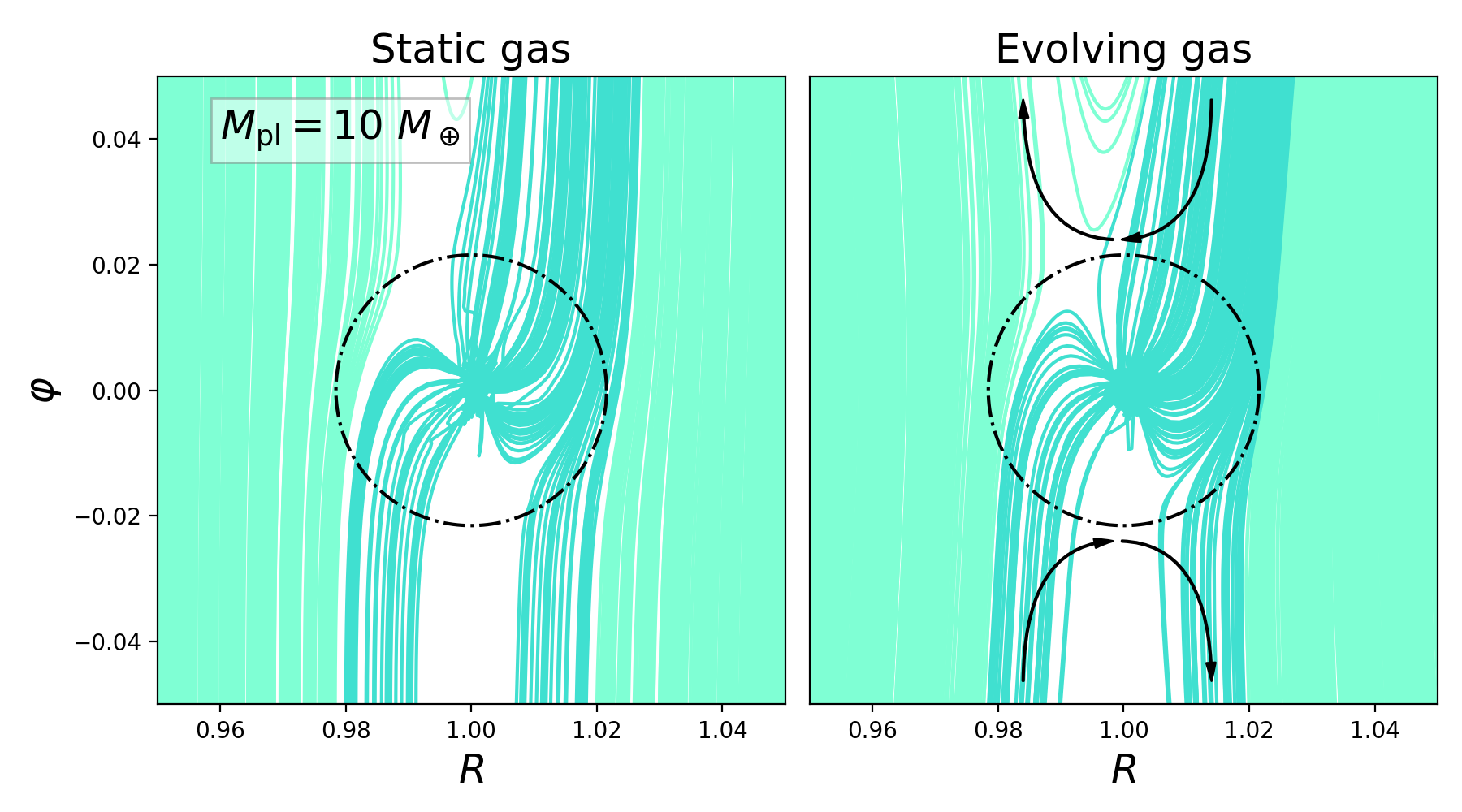}
    \includegraphics[width=0.497\textwidth]{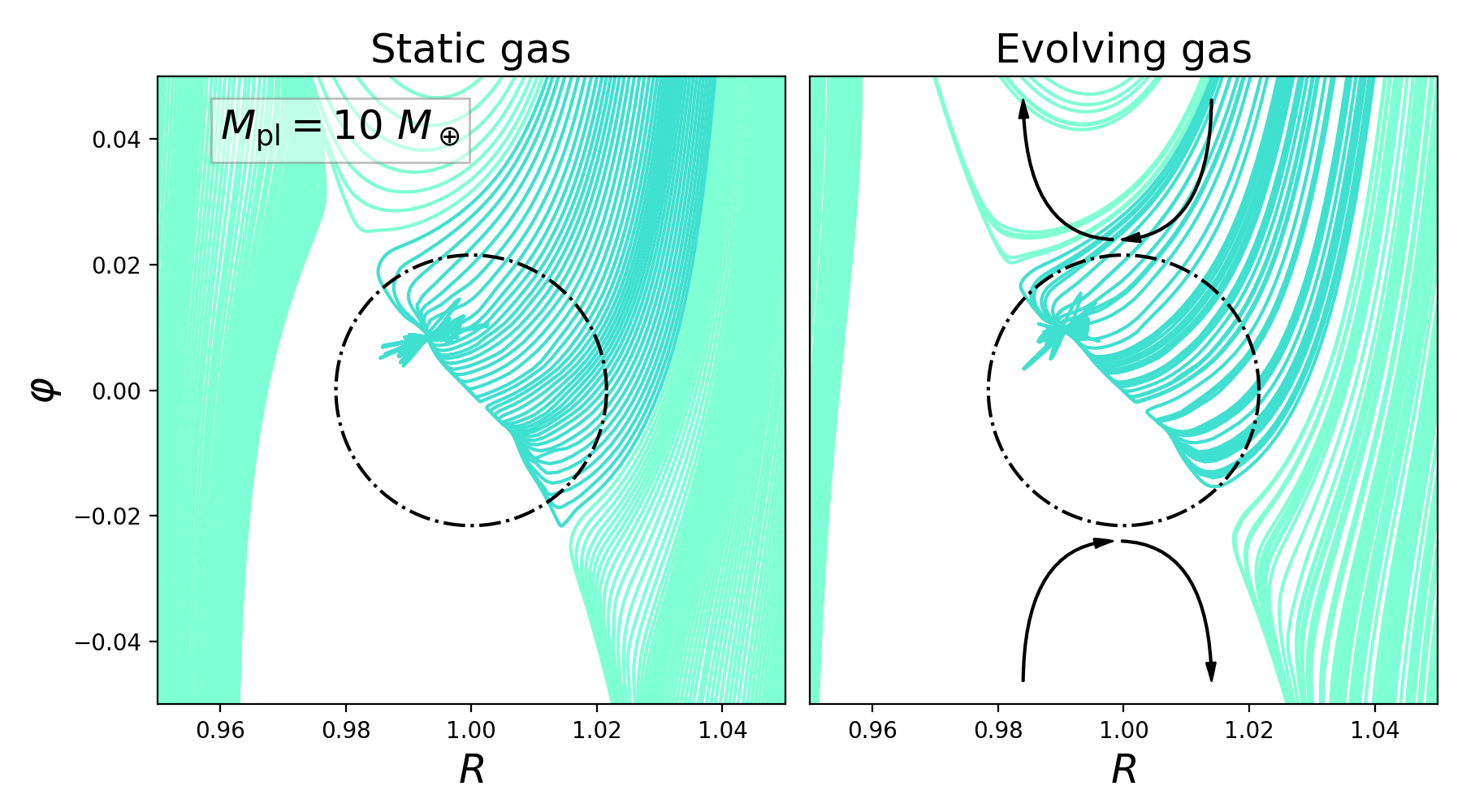}
    \caption{Two monodisperse setups for St = 0.1 (left) and St = 0.889 (right). Top five rows: Three different planet masses (1.5, 4.7, and 10 $M_\oplus$), denoted by colour. The dotted lines signify a static (as opposed to evolving) gas disc. The planet is situated at $a_\mathrm{pl}=1$ AU. Back-reaction is disabled for these results. \textit{First row:} Azimuthally averaged gas surface density. \textit{Second row:} Azimuthally averaged pebble surface density. \textit{Third row:} Total radial pebble mass flow, as described in Eq. \ref{eq:mdot_disc}. \textit{Fourth row:} Total radial gas flow. \textit{Fifth row:} Dimensionless $\eta$-parameter relating $v_{\mathrm{g},\varphi}$ and $v_\mathrm{K}$, as in Eqs. \ref{eq:vg_vk} and \ref{eq:eta}. \textit{Bottom plots:} Pebble streamlines for a static (left), and an evolving disc (right) for a 10 $M_\oplus$ planet. These pebble lines are indicated by the grey lines in the azimuthally averaged and/or summed graphs above. The dash-dotted circle indicates the Hill sphere. The arrows on the right plots illustrating the evolving gas disc signify the direction of the horseshoe orbits. Pebble streamlines arise in the outer disc, with the darker ones accreting onto the planet. The accreting fraction for an evolving gas disc is higher by approximately 16\% for St = 0.1 and lower by approximately 12\% for St = 0.889. }
    \label{fig:difference_sta_evo}
\end{figure*}

We find that gas evolution actually improves the efficiency for the lower Stokes numbers. This effect is more pronounced for higher planetary mass. Contrary to the lower Stokes numbers, the efficiency decreases at the highest Stokes numbers for an evolving disc. This is illustrated in Fig. \ref{fig:difference_sta_evo}, where we compare static and evolving discs. The figure shows results for two Stokes numbers, $\mathrm{St}=0.1$ (left) and $\mathrm{St}=0.889$ (right), each with six simulations: three planet masses (1.5, 4.7, and 10 $M_\oplus$), for both perturbed and unperturbed gas discs. In the first five rows, we show from top to bottom the azimuthally averaged gas surface density $\Sigma_\mathrm{gas}$, pebble surface density $\Sigma_\mathrm{pebbles}$, radial mass flow $\dot{M}_\mathrm{rad}$ for both pebbles and gas, as well as the dimensionless $\eta$-parameter (Eqs. \ref{eq:vg_vk}-\ref{eq:eta}). The latter was easily calculated using an isothermal setup, and the density and local sound speed are direct outputs of FARGO3D. We subsequently calculated $\eta$ via Eq. \ref{eq:eta}, where we used a finite-difference derivative for the pressure $P$. The colours denote planet mass, while the dotted lines show the static disc simulations. 

\begin{table*}[]
\centering
\caption{Fiducial model parameters.}
\resizebox{\textwidth}{!}{%
\begin{tabular}{llllllllllll}
\hline
Dimensions ($r,\varphi$) & Dusty fluids & $M_\star\ \left[M_\odot\right]$  & $\Sigma_\mathrm{g}(r)\ \left[M_\odot \mathrm{AU}^{-2}\right]$ & $M_\mathrm{pl}\ [M_\oplus]$ & $a_\mathrm{pl}\ [\mathrm{AU}]$ & $H_\mathrm{g}/r$   & $f_\mathrm{s/g}$ & Gas perturbed & Back-reaction & $\alpha$ & Diffusion \\ \hline
\texttt{400x2512} & $7$    & $1.0$ & $10^{-4} \left(\frac{r}{1\mathrm{AU}}\right)^{-1/2}$ & $10$ & $1.0$  & $0.05$ & $0.01$          & yes           & yes & $10^{-3}$ & yes           \\ \hline
\end{tabular}%
}
\tablefoot{Note that we adopt code units with $M_\star=1, a_\mathrm{pl}=1, t=\left(GM_\star/a_\mathrm{pl}^3\right)^{-1/2}=1$. Values are shown for a 1 $M_\odot$ with a planet orbiting at 1 AU.}
\label{table:fiducial_model}
\end{table*}

The bottom row with four streamline plots only show the 10 $M_\oplus$ simulations, where the difference between the static and evolving discs are illustrated by the trajectories of the pebbles. The black arrows in the evolving gas plots denote the direction of the horseshoe orbits. Since the static gas disc does not feel the planet, there are no horseshoe orbits there.

For the lowest simulated planet mass (1.51 $M_\oplus$), the inclusion of an evolving gas disc has a negligible effect on all plotted quantities. This is expected, as such a low-mass planet does not significantly perturb the surrounding gas (i.e. for our simulations of gas-only discs, the isolation mass\footnote{For a gas-only disc, we calculated the isolation mass as the mass for which $\eta<0$, that is the point at which inward drifting pebbles halt \citep{Lambrechts_et_al_2014,Bitsch_2018}.} lies between 15 and 17.5 $M_\oplus$). However, at 4.70 $M_\oplus$, still well below the pebble isolation mass, the influence of the planet on the disc becomes apparent. A slight deviation is visible in the gas density (top row), accompanied by a modest change in the dust surface density (second row), a measurable change in the $\eta$-parameter (fifth row), and a slight change in mass flow at $r<a_\mathrm{pl}$ (corresponding to a differing accretion efficiency of the planet).

The latter might be the most interesting effect. For $\mathrm{St} = 0.1$, we see that the change in mass flow, which is equal to the accretion on the planet (Eq. \ref{eq:acc_eff_epsilon}), is higher for the evolving disc compared to the static disc. The planet accretes more due to the gas disc. For higher St, however, we see the opposite effect: the evolving disc decreases the pebble accretion rate. This effect is small (of the order of $\sim10\%$) but measurable.

We believe this effect to be -- at least partly -- qualitatively due to the horseshoe orbits created by the planets gravity in the gas \citep{Murray_Dermott_1999}. Overplotted in the bottom streamline plots of Fig. \ref{fig:difference_sta_evo}, we illustrate those horseshoe orbits schematically. For the high Stokes numbers, we see that the pebbles on the edge of accretion are pushed outwards, deflecting away from the planet and thus decreasing efficiency. Whereas for the lower Stokes, we see the opposite effect. Pebbles on their respective way back for a second turn (bottom left of the plot) are pushed inwards towards the accreting planet, increasing the efficiency.

Our findings -- decreasing efficiency for high St but increasing efficiency for low St -- contradicts the results of \cite{Kuwahara_2020a, Kuwahara_2020b}, who find low St to be less efficient for perturbed discs. But we note these are different findings. First, our lowest simulated St is 0.011, whereas their low St are $<10^{-3}$. Second, they simulate 3D, local, inviscid discs, whereas our simulations are global, 2D, and for a viscous disc ($\alpha=10^{-3}$). Therefore, the two studies are hard to compare.

\subsection{Negligible effect of back-reaction}
\label{subsec:Mono_feedback}

We assessed the impact of including dust back-reaction onto the gas and found its effect on the pebble accretion efficiency to be negligible compared to the evolving disc with no back-reaction. The primary impact of back-reaction is rather seen in the gas dynamics, where we find that the flow reverses direction for discs with larger grains ($\mathrm{St}\gtrsim0.3$), consistent with the results of \cite{Dipierro_2018}. Within the explored planet-mass range for this work, however, the accretion rate and efficiency remain approximately unchanged, and we therefore omit back-reaction from the previous section on monodisperse evolving discs.

At larger planet masses ($M_\mathrm{pl}\gtrsim17.5\ M_\oplus$), back-reaction begins to impact dust dynamics as well, as the solid-to-gas ratio $f_\mathrm{s/g}$ reaches near-unity values. We did incorporate back-reaction, as discussed below, for a polydisperse disc, as it is essential in this case. Without it we would practically simulate multiple monodisperse simulations rather than the coupled polydisperse system we seek to investigate.

\subsection{Polydisperse: A fiducial model}
\label{subsec:Polydisperse}

\begin{figure}[]
	\includegraphics[width=\columnwidth]{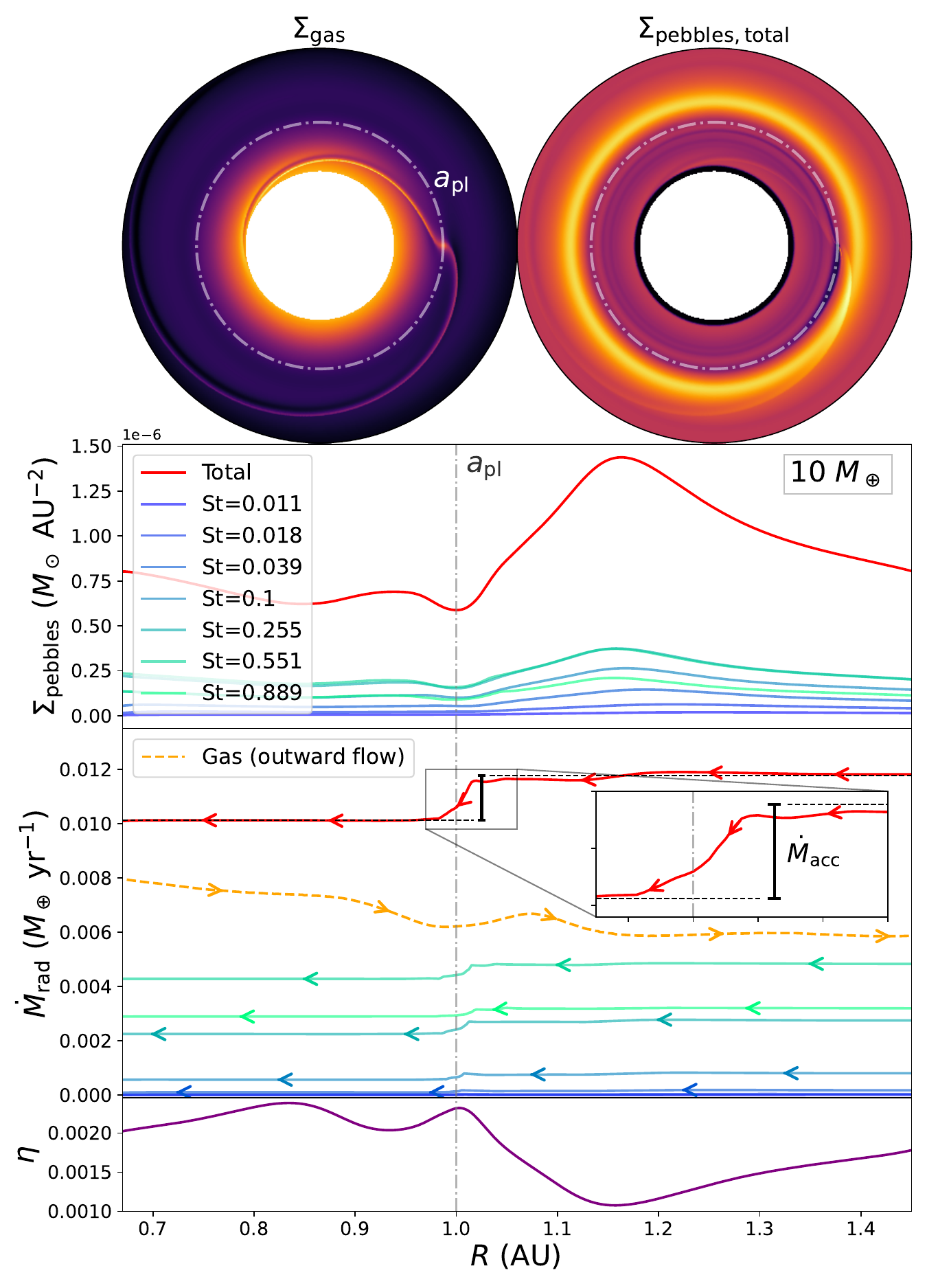}
	\caption{Snapshot after 6000 orbits of the fiducial simulation (Table \ref{table:fiducial_model}) implementing our multi-fluid pebble accretion mechanism. \textit{Top left:} Gas density of the disc. \textit{Top right:} Total pebble density. \textit{First row:} Azimuthally averaged surface density $\Sigma$ for the different dust species. \textit{Second row:} Radial mass flow $\dot{M}_\mathrm{rad}$ of the gas and the different dust species, as explained in Eq. \ref{eq:mdot_disc}. Note that the gas flows outwards due to the back-reaction of the inward moving pebbles. \textit{Third row:} Dimensionless $\eta$-parameter, as explained in Eqs. \ref{eq:vg_vk}-\ref{eq:eta}, relating $v_{\mathrm{g},\varphi}$ with $v_\mathrm{K}$.}
	\label{fig:massflow_example}
\end{figure}

For our polydisperse simulations, we used a fiducial model similar to the monodisperse model previously used. We included the back-reaction of the pebbles on the gas, as well as seven pebble 'fluids' representing a continuous distribution. The size bins and corresponding weights were carefully chosen with the method described in Sect. \ref{subsec:GL_Conjecture}. The values for the chosen disc parameters are shown in Table \ref{table:fiducial_model}. This fiducial simulation and the simulations hereafter were run for many orbits, until the disc reached a stationary structure.

This fiducial model replicates 2D Hill accretion, which interests us for multiple reasons. First, it enables 2D simulations, which greatly reduce computation time. Second, this regime marks the onset of the planet perturbing the gas disc, which is not captured in the classical particle-approach framework (e.g. \citealt{Liu_Ormel_2018}) and assumes an unperturbed gas disc. Studying this regime allows us to investigate the gradual decline of pebble accretion and the associated disc response in a polydisperse setting.

The result of our fiducial model is plotted in Fig. \ref{fig:massflow_example}. On top, we plot the surface density of the gas as well as the pebbles. The first row shows the azimuthally averaged surface density of the pebbles in red. Also plotted are the individual pebble species with nodes $\mathrm{St}_n$ where we assume $\sigma({\rm St}) = \sum_n \sigma_n\delta({\rm St}-{\rm St}_n)$. Already we see a pile-up of pebbles outside the planet similar to the 10 $M_\oplus$ planets in Fig. \ref{fig:difference_sta_evo}. Pebble density increases at this location, such that the solid-to-gas ratio in the pile-up increases to $f_\mathrm{s/g}\approx0.016$.

The second panel shows the radial mass flow of all pebble species (Eq. \ref{eq:mdot_disc}). Also included here is the gas mass flow. Note that the gas moves outwards in contrast to the gas in Fig. \ref{fig:difference_sta_evo}. The reason for this discrepancy is that back-reaction is enabled; the gas reacts to the inward moving mass and, therefore, moves outwards. We note this replicates results of \cite{Dipierro_2018}, who found a reversal of gas radial flow when the mass is dominated by larger, less coupled grains, similar to the distribution in our fiducial model. We highlight the drop in mass flow at $a_\mathrm{pl}$. Similar to the efficiency, this difference is the exact mass accreted by the planet ($\dot{M}_\mathrm{acc}=\Delta\dot{M}_\mathrm{rad}$).

The third panel shows the $\eta$-parameter, relating $v_{\mathrm{g},\varphi}$ and $v_\mathrm{K}$ (Eqs. \ref{eq:vg_vk}-\ref{eq:eta}). We see a dip outside the planet's orbit at the same location where the pebble pile-up is visible (top panel) as well as a small decrease in radial flow (second panel).

\begin{figure}
    \centering
    \includegraphics[width=\columnwidth]{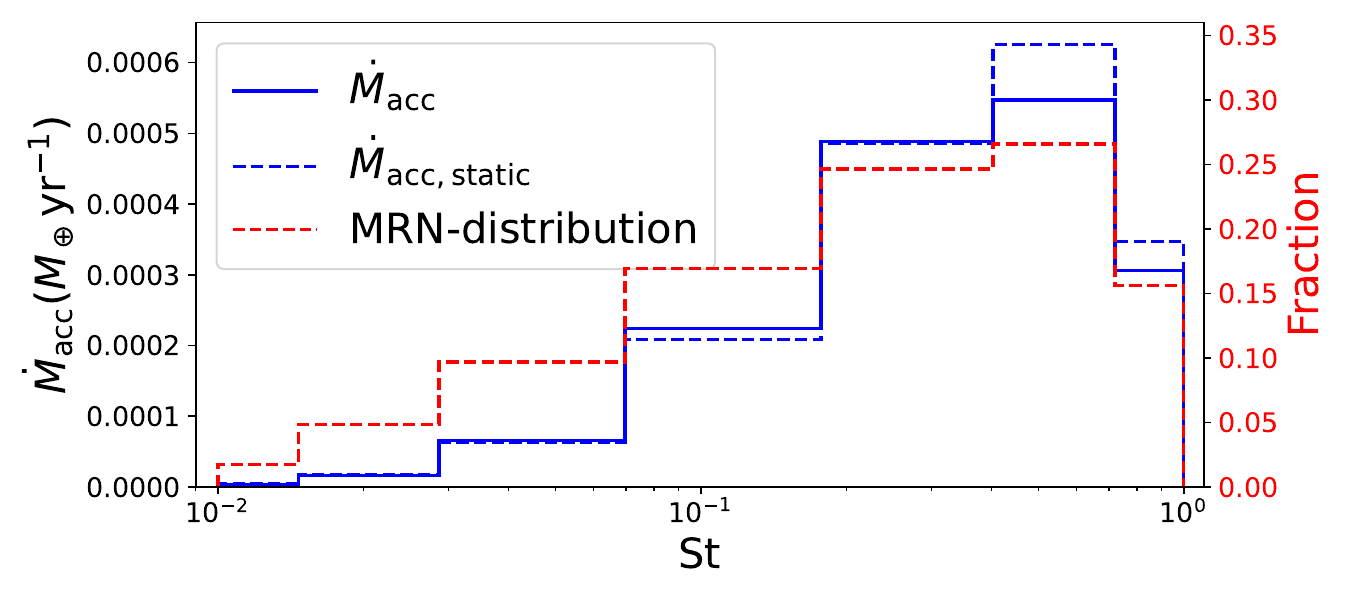}
    \caption{Pebble accretion rate for the 10 $M_\oplus$ planet from Fig. \ref{fig:massflow_example} (blue solid line), and in the static gas disc scenario (blue dashed line), decomposed over the different pebble species. Overlaid in red is the original MRN-distribution, as depicted earlier in Fig. \ref{fig:St_Distribution}.}
    \label{fig:M_acc_decomposed}
\end{figure}

Figure \ref{fig:M_acc_decomposed} shows the total accretion on the planet decomposed over the different pebble species. This is the same $\dot{M}_\mathrm{acc}$ denoted in the second panel, as the difference in $\dot{M}_\mathrm{rad}$ outside and inside the planet signifies the mass accreted onto the planet. The dashed blue lines show the accretion rate for the same simulation for the fiducial model but with a static disc. We see that the larger grains are suppressed by the gas, whereas for the lower St, we observe a small increase due to gas evolution. Overlaid in red is the original MRN-distribution, as seen in Fig. \ref{fig:St_Distribution}. Comparing the two we see the accretion is lob-sided towards higher Stokes numbers. The largest grains are the highest contributors to mass accretion at this stage.

\subsection{Comparison to analytical polydisperse pebble accretion}
\label{subsec:comparison_analyticalPPA}

\begin{figure}
	\includegraphics[width=\columnwidth]{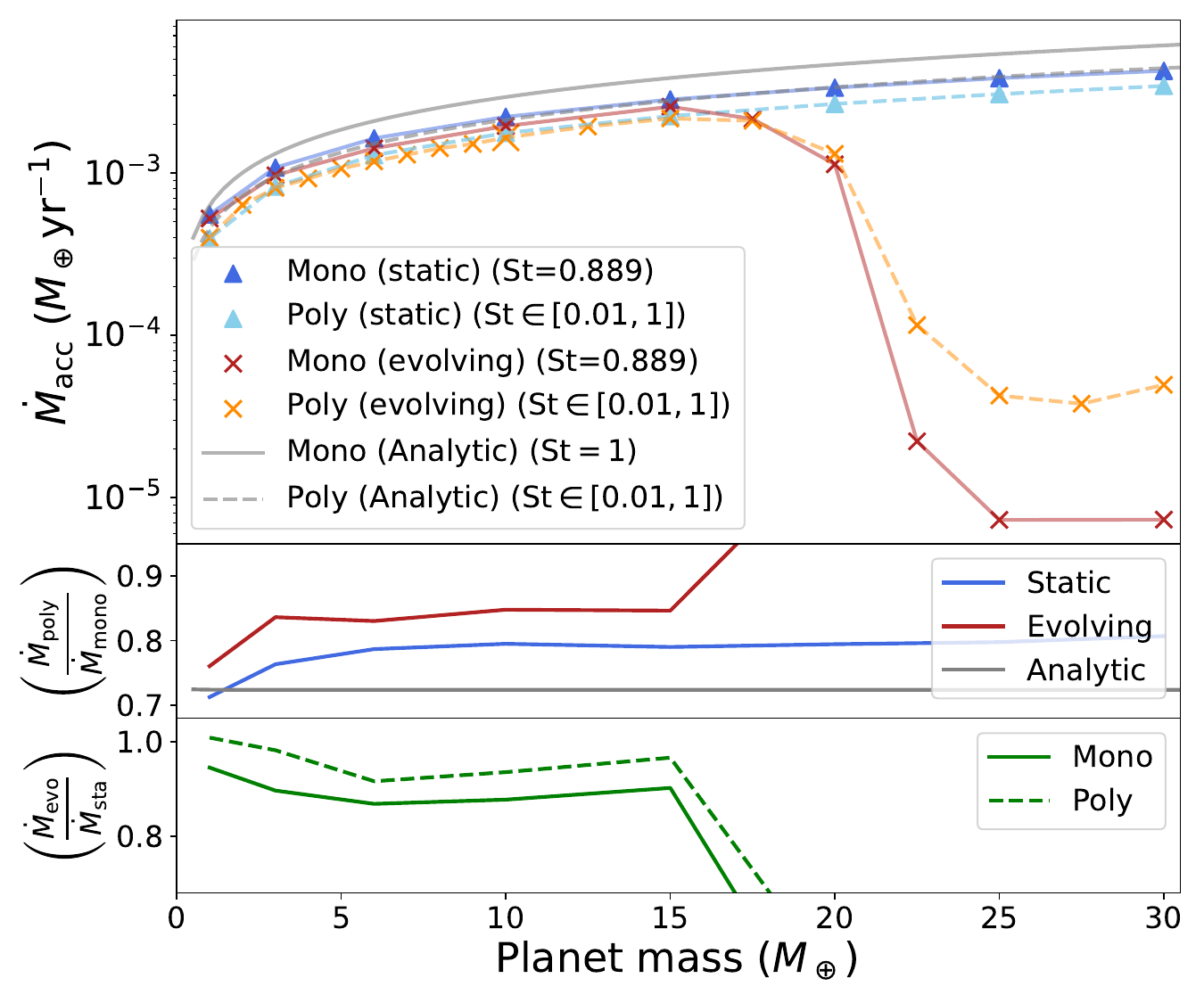}
	\caption{Pebble accretion rates and relative accretion ratios as a function of planet mass. \textit{Top panel:} Mass accretion rate $\dot{M}_\mathrm{acc}$ for different scenarios. The blue triangles and lines represent a static gas disc, while the red and orange crosses and lines indicate the evolving gas disc. The grey lines denote our analytical results, and the solid lines denote the monodisperse populations (St=0.889 for the numerical data; St=1 for the analytic baselines). The enlarged orange cross at $M_\mathrm{pl}=10\ M_\oplus$ denotes the fiducial simulation depicted in Fig. \ref{fig:massflow_example}. \textit{Middle panel:} Ratio between poly- and monodisperse accretion rates, showing the numerical results for a static and evolving disc, as well as the analytic reference for a static disc. \textit{Bottom panel:} Ratio between the evolving and static accretion rates, for the monodisperse (solid) and polydisperse (dashed) populations, highlighting the impact of gas evolution on accretion efficiency. The sharp decline and increase at $M_\mathrm{pl}\gtrsim17.5\ M_\oplus$ in all panels corresponds to the onset of pebble isolation.}
    \label{fig:polydisperse_comparison}
\end{figure}

In Fig. \ref{fig:polydisperse_comparison}, we compare the accretion rates as a function of planet mass obtained from our simulations across three panels. The top panel shows the absolute mass accretion rate, $\dot{M}_\mathrm{acc}$, for different scenarios. 

The solid grey and dashed lines represent the analytical predictions for monodisperse ($\mathrm{St}=0.889$) and polydisperse ($\mathrm{St}\in[0.01,1]$) discs, respectively. These analytical estimations were calculated using the efficiency, $\varepsilon$, taken from \cite{Liu_Ormel_2018}, assuming the same incoming pebble mass flux as in our fiducial disc model (i.e. without the low Stokes approximation). For the polydisperse analytical estimation, we integrated the efficiency over Stokes for an MRN-distribution, where $\mathrm{St}\in[0.01,1]$. 

Our numerical results for static discs (blue symbols) follow the same scaling as the analytical estimations but lie systematically lower. This offset is consistent with the slightly lower efficiencies obtained in our framework for most Stokes numbers compared to the analytical (see Fig \ref{fig:epsilons_monodisperse}).

The numerical results for evolving discs (red and orange symbols) show the same scaling as the static discs, until isolation sets in at $M_\mathrm{pl}\gtrsim17.5\ M_\oplus$. Before isolation, the results for the evolving disc lie slightly lower than the static values. We attribute this to the negative impact of the evolving gas on the higher Stokes numbers (see Fig. \ref{fig:difference_sta_evo}).

To further investigate the impact of the size distribution, the middle panel shows the ratio between the polydisperse and monodisperse accretion rates $\left(\dot{M}_\mathrm{poly}/\dot{M}_\mathrm{mono}\right)$. For a static disc, we find ratios of
\begin{align}
    \label{eq:poly_ratios} \left(\frac{\dot{M}_\mathrm{poly}}{\dot{M}_\mathrm{mono}}\right)_\mathrm{numerical} \approx 0.78 & & \& & &\left(\frac{\dot{M}_\mathrm{poly}}{\dot{M}_\mathrm{mono}}\right)_\mathrm{analytical}\approx0.72.
\end{align}
Notably, both these ratios are significantly higher than 3/7, as found by \cite{Lyra_2023}. As detailed in Appendix \ref{apx:poly_mono_ratio}, this discrepancy arises because previous analytical results rely on the low Stokes approximation -- which underestimates the pebble flux once $\mathrm{St}\gtrsim0.1$ -- since it neglects the $(1+\mathrm{St}^2)$ term from the drift velocity (Eq. \ref{eq:velPert0}).

The bottom panel highlights the relative impact of gas evolution by plotting the ratio between the evolving and static disc accretion rates $\left(\dot{M}_\mathrm{evo}/\dot{M}_\mathrm{sta}\right)$. The monodisperse (solid) line shows a clear decrease, when the ratio is below 1. This reflects our earlier finding that perturbed gas flow decreases the efficiency for the larger grains with $\mathrm{St}\gtrsim0.3$ (see Figs. \ref{fig:efficiency_sta_evo}-\ref{fig:difference_sta_evo}). The polydisperse (dashed) ratio shows values closer to unity. This similarly reflects our findings, as it also contains the smaller grains for which accretion is enhanced by gas perturbations.

Finally, all situations including evolving gas discs illustrate a sharp drop at $M_\mathrm{pl}\gtrsim17.5\ M_\oplus$, marking the onset of pebble isolation \citep[e.g.][]{Lambrechts_et_al_2014, Bitsch_2018}. In this post-isolation regime, the accretion rate for our polydisperse disc is nearly an order of magnitude higher than the monodisperse case. This is because the polydisperse population includes smaller grains that are capable of diffusing across the planet-generated pressure bump, whereas the inward flux of larger monodisperse pebbles is effectively terminated. The exact mechanics of how the pebble isolation mass is influenced by disc parameters is beyond the scope of this study and will addressed in future work.

\section{Discussion}
\label{sec:discussion}

\subsection{(Multi-)Fluid prescription versus particle approach}

Our results demonstrate that pebble accretion can be modelled accurately using a multi-fluid hydrodynamic framework. The accretion efficiencies we obtain are consistent with those from previous particle-based studies, validating the multi-fluid method as a reliable alternative. This agreement confirms that, despite the conceptual differences between Lagrangian particle tracking and Eulerian fluid descriptions, the essential physics of pebble accretion are well captured in a multi-fluid treatment.

An important advantage of the multi-fluid framework emerges at higher planet masses, close to the isolation mass. As the planetary mass increases, the gas disc becomes increasingly perturbed, altering the local pressure gradient and gas velocity field. In particle-based approaches, such perturbations can complicate the interpretation of pebble trajectories and introduce several numerical and physical challenges. Specifically, implementing gas-particle back-reaction in a Lagrangian framework requires mapping the momentum exchange of discrete particles onto the Eulerian gas grid. This often leads to particle shot noise unless an extremely large number of particles is used \citep[e.g.][]{Peirano_2006}. Furthermore, as the planet mass increases and gaps or pile-ups form, particles tend to cluster in high-density regions (e.g. the pressure bump), leaving other parts of the disc undersampled and making it numerically difficult to maintain a converged feedback calculation.

In a polydisperse scenario, these challenges are amplified because each size bin requires a sufficient number of particles to avoid sampling errors, which becomes computationally prohibitive when approximating a continuous distribution. In contrast, the multi-fluid method treats each pebble species as a continuous density field, allowing the momentum exchange and back-reaction to be handled self-consistently within the fluid solver without shot noise, naturally incorporating the feedback between the gas and multiple pebble species.

This makes the multi-fluid approach particularly well suited for studying later stages of growth, where gap formation and strong pressure perturbations modify pebble fluxes. In this regime, the assumption of an unperturbed background disc becomes invalid, and a hydrodynamic treatment becomes essential. Therefore, the framework developed here provides a natural pathway towards investigating pebble isolation mass, where the planet-induced pressure bump halts inward drift. While a detailed study of isolation mass lies beyond the scope of this paper, our results demonstrate that the present method is well positioned to address this problem in future work.

\subsection{Change in pebble accretion efficiency for perturbed gas discs}

Allowing the gas disc to evolve modifies the accretion efficiency in a systematic, near-linear manner (Fig. \ref{fig:efficiency_sta_evo}). For $\mathrm{St} \lesssim 0.3$, the efficiency in the evolving disc is higher than in the static case, whereas for $\mathrm{St} \gtrsim 0.3$ it is reduced. The transition occurs around $\mathrm{St} \sim 0.3$ and becomes more pronounced with increasing planetary mass. This trend is robust across the explored mass range; therefore, we believe it to reflect a structural change in the flow rather than stochastic variability.

We believe this can be explained, at least in part, qualitatively by examining the streamlines of the pebbles in the evolving models compared to the static discs (bottom plots of Fig. \ref{fig:difference_sta_evo}). When the gas is allowed to respond to the planet, a horseshoe region develops that is absent in the static prescription. Particles with low Stokes numbers, which remain tightly coupled to the gas, follow these horseshoe trajectories. During the inward leg of the horseshoe turn, material that has already passed ($a_\mathrm{pl}$) from the outer disc is redirected towards the planet’s vicinity. 
This essentially gives the pebbles a second chance to be accreted. This 'second-chance' mechanism mirrors the behaviour of scattered aerodynamically large pebbles, as found by \cite{Huang_Ormel_2023}. While they focussed on the 3D accretion of larger pebbles ($\mathrm{St}\gg1$), our 2D multi-fluid results suggest a qualitatively similar promotive effect for more tightly coupled grains ($\mathrm{St}\lesssim0.3$).

In contrast, particles with higher Stokes numbers are only partially coupled to the gas and do not complete the same turnaround motion. Instead, they are more readily displaced away from the planet’s feeding region during the horseshoe exchange, reducing the effective accretion cross section. While a fully quantitative description is lacking here (it would require a dedicated analysis of particle trajectories and torque balance), the correlation between the emergence of the horseshoe flow and the sign change in $(\varepsilon_{\rm sta}-\varepsilon_{\rm evo})/\varepsilon_{\rm sta}$ strongly suggests that the modified co-orbital dynamics are responsible for the St-dependent efficiency shift.

\subsection{Polydisperse accretion and its effects}

The polydisperse treatment reveals that pebble accretion is intrinsically biased towards higher Stokes numbers, even more so than the underlying disc distribution (Fig. \ref{fig:M_acc_decomposed}). While the MRN-distribution already favours larger particles in terms of mass content, the accretion process itself amplifies this asymmetry. This amplification is a direct consequence of the scaling of the accretion radius with the Stokes number; in the Hill regime this scaling follows $R_\mathrm{acc}\propto\mathrm{St}^{1/3}$. Consequently, the effective distribution of accreted material becomes more top-heavy than the background population in the disc.

A key result of our polydisperse framework is that the ratio between polydisperse and monodisperse accretion (Eq. \ref{eq:poly_ratios}) is noticeably higher than the 3/7 value previously estimated by \cite{Lyra_2023}. As detailed in Appendix \ref{apx:poly_mono_ratio}, this discrepancy arises because earlier analytical results rely on a low Stokes approximation for the incoming pebble flux. This approximation remains valid for very small grains but fails to capture the correct flux for pebbles with $\mathrm{St}\gtrsim0.1$. Since our distribution includes pebbles up to $\mathrm{St}=1$, the removal of this approximation yields a higher ratio that aligns with our numerical findings.

Finally, the net impact of gas evolution on the total accretion rate is intrinsically linked to the assumed size distribution. In our fiducial model, which adopts an MRN-distribution with $\mathrm{St}\in[10^{-2},1]$, we find that perturbing the gas disc overall reduces the total pebble accretion rate compared to a static disc. As shown in Sect. \ref{subsec:Mono_perturbing}, the perturbed gas flow promotes accretion for small pebbles ($\mathrm{St}\lesssim0.3$) but hinders it for larger pebbles ($\mathrm{St}\gtrsim0.3$). Because the MRN-distribution is mass-dominated by these larger Stokes numbers, their reduced efficiency outweighs the gains made by the smaller, more tightly coupled grains. This suggests that the influence of a planet on its own growth rate is highly sensitive to the pebble population. A disc dominated by significantly smaller grains could potentially show a reverse trend, where gas perturbations cause a net increase in the total mass accretion rate.

\subsection{Limitations and future work}

Several limitations of the present study should be acknowledged. First, our simulations are strictly 2D. While our 2D simulations capture the essential radial and azimuthal dynamics of pebble accretion, they neglect the complex vertical structure of the gas flow around the planet. For low-mass planets, such as those investigated in this study, 3D gas dynamics become essential. Specifically, realistic 3D gas flows are characterised by a strong midplane outflow and complex circulation patterns that are not captured in a vertically integrated 2D framework.

Since small pebbles are more tightly coupled to the gas, their trajectories are heavily influenced by these 3D flow patterns. The characteristic midplane outflow could significantly modify the accretion efficiency for these small grains, potentially altering our conclusions regarding the 'promotive' effect of the horseshoes (see Fig. \ref{fig:difference_sta_evo}). Extending this framework to three dimensions is a next step to fully quantify the accretion rates for low-mass protoplanets. This would enable the simultaneous treatment of both the 3D gas flow structures, the vertical settling, and the stratification of the various pebble species.

Second, turbulent diffusion was implemented in a simplified manner. Although this is sufficient for the parameter space explored here, a more detailed turbulence model may become important when studying pebble isolation or gap formation, where subtle changes in pressure gradients can determine whether particles are trapped or continue drifting inwards.

A particularly promising avenue for future work is the study of pebble isolation mass within the multi-fluid framework. Because the method tracks multiple pebble species simultaneously, all coupled to the gas and therefore each other, it provides direct insight into how different Stokes numbers respond to the formation of a pressure bump as well as how the pressure bump reacts to the subsequent dust traffic jam. This allows one to determine not only when accretion halts, but also which size ranges are most efficiently filtered. Such a self-consistent treatment is difficult to achieve in simplified or monodisperse models.

\section{Conclusions}
\label{sec:conclusions}
\begin{itemize}
    \item We demonstrate that polydisperse pebble accretion can be accurately modelled using a multi-fluid hydrodynamic framework. This approach yields results consistent with classical particle-based studies, provided that the planetary mass is sufficient for the accretion radius $R_\mathrm{acc}$ to be well-resolved by the numerical grid cells of the dust fluid.
	\item Perturbing the gas disc modifies the accretion efficiency. For $\mathrm{St}\lesssim0.3$, the efficiency increases relative to the static-disc case, whereas for $\mathrm{St}\gtrsim0.3$ it decreases. This trend becomes more pronounced with increasing planetary mass. We emphasize that this conclusion is based on 2D simulations. A realistic 3D gas flow would significantly influence the pebble accretion efficiency for small pebbles.
	\item In our assumption of an MRN-distribution with $\mathrm{St}\in[10^{-2},1]$, evolving the gas disc lowers the total accretion rate, since the mass distribution is dominated by pebbles with $\mathrm{St}\gtrsim0.3$.
	\item We find that back-reaction remains negligible for the planet masses studied here ($\leq10M_\oplus$), as the solid-to-gas mass ratio $f_\mathrm{s/g}$ never exceeds $\sim0.016$. Consequently, the accretion rate differs only by a few percent.
    \item Pebble accretion in a polydisperse disc with an MRN-distribution is intrinsically biased towards higher Stokes numbers. This bias arises from two factors: first, the MRN-distribution is inherently mass-dominated by larger grains, and second, the accretion process itself amplifies this asymmetry because the accretion radius scales with the Stokes number, providing larger pebbles with a significantly greater accretion cross section.
    \item We find the ratio between the poly- and monodisperse accretion rates to increase noticeably above previous estimations of $\left(\dot{M}_\mathrm{poly}/\dot{M}_\mathrm{mono}\right)=3/7$. This increase occurs once $\mathrm{St}_\mathrm{max}\gtrsim0.1$. For this ratio we find an analytical value of $\sim0.72$ and a value of $\sim0.78$ for our simulations, both for an MRN-distribution where $\mathrm{St}\in[10^{-2},1]$.
\end{itemize}

\section*{Data availability}

The data that support the findings of this study are openly available in 4TU.ResearchData under the name: 'Repository supporting the publication: A multi-fluid approach for pebble accretion' at \href{doi.org/10.4121/4f262c6a-cee6-4d5f-90de-57a217450c37}{doi.org/10.4121/4f262c6a-cee6-4d5f-90de-57a217450c37}


\begin{acknowledgements}
	The authors thank the anonymous referee for helpful suggestions that greatly improved this manuscript. We also thank Jip Matthijsse, Benjamin Silk, and Hossam Aly for their input and helpful discussions. TJK further thanks Michiel Lambrechts, Anders Johansen, Ayumu Kuwahara, Wladimir Lyra \& Satoshi Okuzumi for their fruitful discussions while on visit in Copenhagen. The authors acknowledge the use of computational resources of the DelftBlue supercomputer, provided by \cite{Delftblue}. This project has received funding from the European Research Council (ERC) under the European Union’s Horizon Europe research and innovation programme (Grant Agreement No. 101054502). This work made use of several open-source software packages. We acknowledge FARGO3D \citep{FARGO3D}, numpy \citep{Numpy}, matplotlib \citep{Matplotlib}, and scipy \citep{Scipy}.
\end{acknowledgements}


\bibliographystyle{aa}
\bibliography{references}


\begin{appendix}

\section{Dust streamline calculations}
\label{apx:Velocityfield}

The velocity fields in Figs. \ref{fig:vel_fields} \& \ref{fig:difference_sta_evo} use the linear perturbation theory. The derivation of how this is done is shown below.

We start from the dust momentum conservation Eq. \ref{eq:momentumconvdust}, in the monodisperse limit:
\begin{align}
	\partial_t\mathbf{v} + \left(\mathbf{v}\cdot\nabla\right)\mathbf{v} =& -\nabla\Phi_* - \frac{\mathbf{v} - \mathbf{v}_\mathrm{K}}{\tau_\mathrm{s}} + 
    \frac{\mathbf{v}_{\rm g} - \mathbf{v}_\mathrm{K}}{\tau_\mathrm{s}} -\nabla\Phi_{\rm p}\nonumber\\
    \equiv & -\nabla\Phi_* - \frac{\mathbf{v} - \mathbf{v}_\mathrm{K}}{\tau_\mathrm{s}} + 
    \Pi' -\nabla\Phi',
\end{align} 
where we omitted the subscript 'd' for clarity. The gas disc is taken to be static, with a purely azimuthal velocity $v_{{\rm g},\varphi}=(1-\eta)v_{\rm K}$. For $\eta \ll {\rm St}$, and a sufficiently small planet mass, we can treat $\Pi'$ and $\Phi'$ as small perturbations. It is straightforward to see that when $\Pi'=\Phi'=0$, the equilibrium velocity field is ${\bf v} = {\bf v}_{\rm K}$.

We can describe linearised equations in terms of the angular momentum perturbation $L'=rv_\varphi'$:
\begin{align}
	\label{eq:MomPertR} \left(\partial_t + \frac{v_\varphi}{r}\partial_\varphi\right)v_r' - \frac{2v_\varphi}{r^2}L' &= -\partial_r\Phi' -\frac{v_r'}{\tau_\mathrm{s}},\\
	\label{eq:MomPertPhi} \left(\partial_t + \frac{v_\varphi}{r}\partial_\varphi\right)L'  + \frac{v_\varphi}{2}v_r' &= r\Pi' - \partial_\varphi\Phi' - \frac{L'}{\tau_\mathrm{s}}.
\end{align}

\subsection{Fourier decomposition}
\label{subsec:fourier}

As is standard in disc-planet interactions \citep{GT80}, we assume the perturbations to be of the form:
\begin{equation}
	X'(r,\varphi,t) = \hat{X}_m(r) \exp{\left(im\varphi - im\Omega_0t\right)},
\end{equation}
where $m$ is the (integer) azimuthal wave number and $\Omega_0$ is the angular velocity of the planet. By choosing this form of a perturbation, they are steady in the frame co-rotating with the planet. Implementing this in Eqs. \ref{eq:MomPertR}-\ref{eq:MomPertPhi}:
\begin{align}
	\sigma\hat{v}_m - \frac{2v_\varphi}{r^2}\hat{L}_m &= -\partial_r\hat{\Phi}_m,\\
	\sigma\hat{L}_m + \frac{v_\varphi}{2	}\hat{v}_m &= \delta_{m0}r\hat{\Pi} - im\hat{\Phi}_m,
\end{align}
where $\sigma\equiv im\left(\Omega - \Omega_0\right)+1/\tau_\mathrm{s}$ and $\Omega=v_{\varphi}/r$ denotes the angular velocity. This system can be straightforwardly solved:
\begin{align}
	\hat{v}_m &= -\frac{1}{r}\frac{2 \Omega i m \hat{\Phi}_m - 2\Omega \delta_{m0}r\hat{\Pi} + \sigma r \partial_r \hat{\Phi}_m}{\sigma^2 + \Omega^2},\\
	\hat{L}_m &= -\frac{\sigma i m \hat{\Phi}_m - \sigma \delta_{m0} r \hat{\Pi} + \frac{1}{2} \Omega r \partial_r\hat{\Phi}_m}{\sigma^2 + \Omega^2}.
\end{align}
From here we can describe the total velocity and angular momentum perturbations as:
\begin{align}
	v_r' &= \sum_{m=0}^{\infty}\hat{v}_m(r)\exp{\left(im\varphi - im\Omega_0t\right)},\\
	L' &= \sum_{m=0}^{\infty}\hat{L}_m(r)\exp{\left(im\varphi - im\Omega_0t\right)}.
\end{align}
In the absence of a planet, we only retain $m=0$ and find:
\begin{align}
	v_r' &= \frac{2r\Omega\tau_\mathrm{s}\left(\Omega_\mathrm{g} - \Omega_\mathrm{K}\right)}{1 + \Omega^2\tau_\mathrm{s}^2},\\
	\Omega' &= \frac{\Omega_\mathrm{g} - \Omega_\mathrm{K}}{1 + \Omega^2\tau_\mathrm{s}^2}.
\end{align}
These describe the usual radial and azimuthal dust drift. For $\Phi' \neq 0$, the number of terms required in the Fourier series depends on the smoothing length; smaller values of $\lambda$ require additional terms to achieve a converged solution. 

\subsection{Series in Stokes number}

For dust particles with Stokes number ${\rm St} = \Omega \tau_{\rm s} \ll 1$, we can use an alternative to the Fourier decomposition, which can be an advantage for very small smoothing lengths. Combine the steady state (in the frame corotating with the planet) versions of Eqs. \ref{eq:MomPertR} and \ref{eq:MomPertPhi}:
\begin{align}
  (\Omega-\Omega_0)^2\partial_\varphi^2 u' +
  \frac{2(\Omega-\Omega_0)}{\tau}\partial_\varphi u'
  + u'\left(\Omega^2+\frac{1}{\tau^2}\right) =\nonumber\\ 2\Omega \Pi'-\frac{2\Omega}{r}\partial_\varphi\Phi' - \frac{\partial_r\Phi'}{\tau} -
  (\Omega-\Omega_0)\partial_{r,\varphi}^2 \Phi' .    
\end{align}
Since the only derivatives of the unknown function $u'$ are with respect to $\varphi$, we can treat this as an ODE. While it is possible to write down the general solution, it is of not much use in practice. Assuming the Stokes number to be constant, we can however develop a series in ${\rm St}\ll 1$, $u'=\sum_n u_n' {\rm St}^n$, with
\begin{align}
  u'_0 =& 0,\\
  u'_1 =& -2\eta r\Omega - \frac{\partial_r\Phi'}{\Omega},\\
  u'_2 =& -\frac{2}{r\Omega}\partial_\varphi\Phi' +
          (\Omega-\Omega_0)\frac{\partial_{r,\phi}^2 \Phi'}{\Omega^2},\\
 u'_n =& -u'_{n-2} - \frac{2(\Omega-\Omega_0)}{\Omega}\partial_\varphi u'_{n-1} - \frac{(\Omega-\Omega_0)^2}{\Omega^2}\partial_\varphi^2 u'_{n-2},\\
  v'_0 =& -\eta r\Omega,\\
  v'_1 =& -\frac{1}{r\Omega}\partial_\varphi\Phi',\\
  v'_n =&- \frac{u'_{n-1}}{2} - \frac{(\Omega-\Omega_0)^2}{2\Omega^2}\partial_\varphi^2 u'_{n-1}.
\end{align}

\begin{figure}
	\includegraphics[width=\columnwidth]{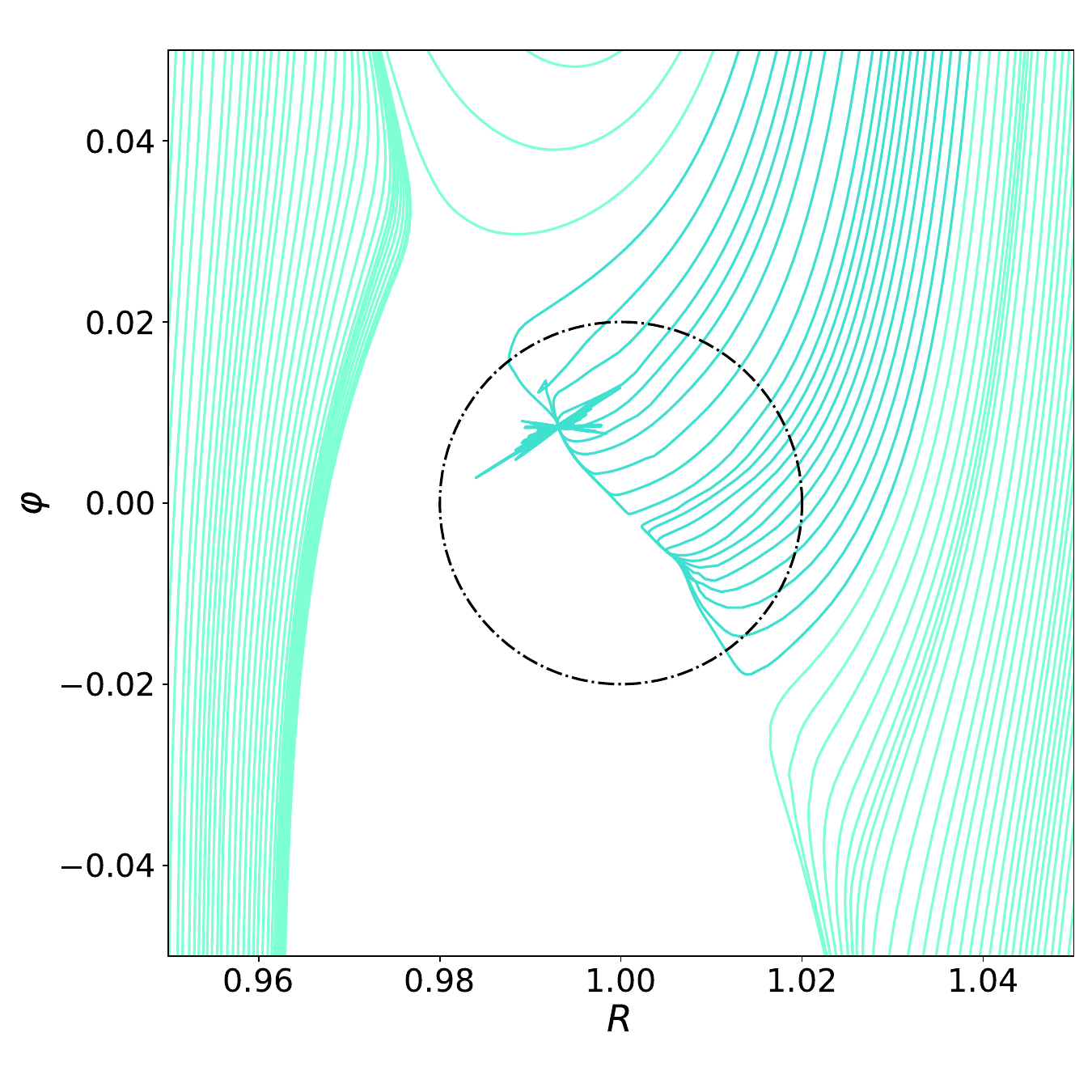}
	\caption{Example of streamlines created using this method. Illustrated is a local co-rating shearing box of a 10 $M_\oplus$ planet.}
    \label{fig:apx_streamlines}
\end{figure}

See Fig. \ref{fig:apx_streamlines} as an example for these calculated streamlines.

\section{Analytical ratio between polydisperse and monodisperse accretion rate}
\label{apx:poly_mono_ratio}
We find the (analytical) ratio between poly- and monodisperse accretion to be $\sim0.72$ (Eq. \ref{eq:poly_ratios}), while previous analytical results found an exact factor of 3/7 for this ratio \citep{Lyra_2023}. This appendix gives a short explanation of why we expect a higher ratio, due only to the low Stokes number approximation used previously.

The monodisperse accretion rate can be written as the total radial mass flow of pebbles times the accretion efficiency:
\begin{equation}
    \dot{M}_\mathrm{mono} = 2\pi r v_r \Sigma_\mathrm{p}\varepsilon(\mathrm{St}_\mathrm{mono}),
\end{equation}
where $v_r$ is the radial drifting velocity of the pebbles:
\begin{equation}
    v_r = \frac{2\mathrm{St}}{1+\mathrm{St}^2}\eta v_\mathrm{K},
\end{equation}
which is the usual drift solution \citep{Weidenschilling_1977a}, similar to Eq. \ref{eq:velPert0}. 
We can calculate the polydisperse accretion rate in a similar manner. We just take the size density $\sigma(\mathrm{St})$ defined in Eq. \ref{eq:size_density} to find:
\begin{equation}
	\dot{M}_\mathrm{poly}=4\pi r\eta v_\mathrm{K} \int_{\mathrm{St}_\mathrm{min}}^{\mathrm{St}_\mathrm{max}}\frac{\sigma(\mathrm{St})\mathrm{St}\varepsilon(\mathrm{St})}{1 + \mathrm{St}^2}\mathrm{dSt}.
\end{equation}
The ratio between these two accretion rates is then:
\begin{equation}
	\frac{\dot{M}_\mathrm{poly}}{\dot{M}_\mathrm{mono}}=\frac{1}{\Sigma_\mathrm{p}}\frac{1 + \mathrm{St}_\mathrm{mono}^2}{\mathrm{St}_\mathrm{mono}\varepsilon(\mathrm{St}_\mathrm{mono})} \int_{\mathrm{St}_\mathrm{min}}^{\mathrm{St}_\mathrm{max}} \frac{\sigma(\mathrm{St})\mathrm{St}\varepsilon(\mathrm{St})}{1 + \mathrm{St}^2}\mathrm{dSt}.
\end{equation}
We assume $\varepsilon$ and $\sigma$ to be power laws, so we can take $\varepsilon\propto\mathrm{St}^{a}$ and $\sigma\propto\mathrm{St}^{b}$ to find:
\begin{equation}
	\frac{\dot{M}_\mathrm{poly}}{\dot{M}_\mathrm{mono}}=\frac{1+b}{\mathrm{St}^{b+1}_\mathrm{max}-\mathrm{St}^{b+1}_\mathrm{min}}\frac{1+\mathrm{St}_\mathrm{mono}^2}{\mathrm{St}_\mathrm{mono}^{1+a}}\int_{\mathrm{St}_\mathrm{min}}^{\mathrm{St}_\mathrm{max}} \frac{\mathrm{St}^{1+a+b}}{1+\mathrm{St}^2}\mathrm{dSt},
\end{equation}
since $\Sigma_\mathrm{P}=\int\sigma(\mathrm{St})\mathrm{dSt}$.

In our assumption for an MRN-distribution ($b=-1/2$) and taking the St-dependency for the efficiency from \cite{Liu_Ormel_2018} and \cite{Lambrechts_Johansen_2014} ($a=-1/3$), this means:
\begin{equation}
	\label{eq:apx_analytical_accrate_ratio} \frac{\dot{M}_\mathrm{poly}}{\dot{M}_\mathrm{mono}}=\frac{1/2}{\mathrm{St}^{1/2}_\mathrm{max}-\mathrm{St}^{1/2}_\mathrm{min}}\frac{1+\mathrm{St}_\mathrm{max}^2}{\mathrm{St}_\mathrm{max}^{2/3}}\int_{\mathrm{St}_\mathrm{min}}^{\mathrm{St}_\mathrm{max}} \frac{\mathrm{St}^{1/6}}{1+\mathrm{St}^2}\mathrm{dSt},
\end{equation}
assuming $\mathrm{St}_\mathrm{mono}=\mathrm{St}_\mathrm{max}$. 

For $[\mathrm{St}_\mathrm{min}, \mathrm{St}_\mathrm{max}]=[0,1]$, this ratio would be approximately $\sim0.66$. Our result of Eq. \ref{eq:poly_ratios} is higher, because we have a finite $\mathrm{St}_\mathrm{min}$. If we put in our distribution of $\mathrm{St}\in[10^{-2},1]$, the ratio becomes $\sim0.72$, as is also portrayed by the grey line in the middle panel of Fig. \ref{fig:polydisperse_comparison}.

If we use the small Stokes approximation thereby neglecting the $(1 + \mathrm{St^2})$ term from Eq. \ref{eq:apx_analytical_accrate_ratio} we find:
\begin{align}
    \nonumber \frac{\dot{M}_\mathrm{poly}}{\dot{M}_\mathrm{mono}} &= \frac{1/2}{\mathrm{St}^{1/2}_\mathrm{max}-\mathrm{St}^{1/2}_\mathrm{min}}\frac{1}{\mathrm{St}_\mathrm{max}^{2/3}}\int_{\mathrm{St}_\mathrm{min}}^{\mathrm{St}_\mathrm{max}} \mathrm{St}^{1/6}\mathrm{dSt},\\
    &=\frac{3}{7} \frac{\mathrm{St}_\mathrm{max}^{7/6} - \mathrm{St}_\mathrm{min}^{7/6}}{\mathrm{St}^{1/2}_\mathrm{max}-\mathrm{St}^{1/2}_\mathrm{min}}\frac{1}{\mathrm{St}_\mathrm{max}^{2/3}}.
\end{align}
Which for $\mathrm{St}_\mathrm{min}=0$ is exactly 3/7, the original solution of \cite{Lyra_2023}.

\begin{figure}
    \centering
    \includegraphics[width=\columnwidth]{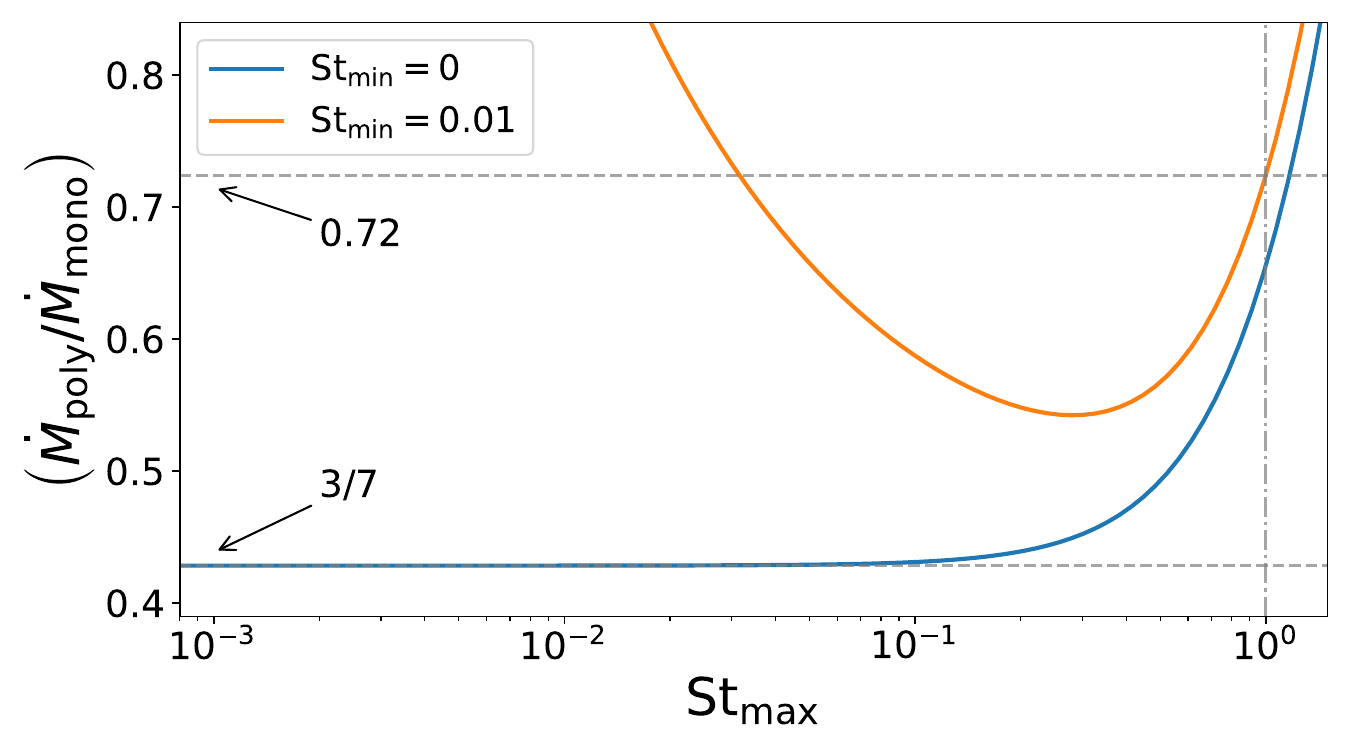}
    \caption{Ratio between the poly- and monodisperse accretion rates (Eq. \ref{eq:apx_analytical_accrate_ratio}) for two different $\mathrm{St}_\mathrm{min}$. The horizontal dashed lines signify $\left(\dot{M}_\mathrm{poly}/\dot{M}_\mathrm{mono}\right)=0.72$, our analytical result from Eq. \ref{eq:poly_ratios}, and $\left(\dot{M}_\mathrm{poly}/\dot{M}_\mathrm{mono}\right)=3/7$, the result of \cite{Lyra_2023}. The vertical dash-dotted line is a visual aid for $\mathrm{St}_\mathrm{max}=1$, our considered maximum value.}
    \label{fig:apx_accratio}
\end{figure}

We plotted the effect of $\mathrm{St}_\mathrm{max}$ on this ratio in Fig. \ref{fig:apx_accratio}. The blue line keeps the theoretical value of $\mathrm{St}_\mathrm{min}=0$, while we plot the orange line with $\mathrm{St}_\mathrm{min}=0.01$. The vertical dash-dotted line is a visual aid of $\mathrm{St}_\mathrm{max}=1$, our maximum value. We also plot two horizontal dashed lines as a visual aid. The top line denotes our analytical value from Eq. \ref{eq:poly_ratios}, while the bottom line denotes the analytical value of 3/7 found by \cite{Lyra_2023}.

\end{appendix}


\end{document}